\PassOptionsToPackage{colorlinks=true,linkcolor=blue,citecolor=blue}{hyperref}
\documentclass[prd,twocolumn,nofootinbib,superscriptaddress,letterpaper]{revtex4}

\newcommand\skipme[1]{}

\newcommand\ForInternalReference[1]{}
\newcommand\SkipForEarlyCirculation[1]{}

\newcommand\SkipPP[1]{}
\usepackage{booktabs}
\usepackage{tabularx}
\usepackage{makecell}
\usepackage{array}
\usepackage{graphicx}
\usepackage{dcolumn}
\usepackage{bm}
\usepackage{xspace}
\usepackage{url}
\usepackage{amsmath}
\usepackage{longtable}
\usepackage{listings}
\usepackage{multirow}
\usepackage{amssymb}
\usepackage{color}
\usepackage{tikz}
\usepackage{orcidlink}
\usepackage{cleveref}
\usepackage{xurl}
\usetikzlibrary{shapes.geometric, arrows, bayesnet}
\usepackage{hyperref}
\newcommand\optional[1]{}

\tikzstyle{startstop} = [circle, rounded corners, minimum width=1cm, minimum height=1cm,text centered, draw=black, fill=red!30]
\tikzstyle{io} = [trapezium, trapezium left angle=70, trapezium right angle=110, minimum width=2cm, minimum height=1cm, text centered, draw=black, fill=blue!30]
\tikzstyle{process} = [rectangle, minimum width=2cm, minimum height=1cm, text centered, draw=black, fill=orange!30]
\tikzstyle{decision} = [diamond, minimum width=2cm, minimum height=1cm, text centered, draw=black, fill=green!30]
\tikzstyle{arrow} = [thick,->,>=stealth]
\definecolor{amber}{rgb}{1.0, 0.75, 0.0}
\definecolor{orange}{rgb}{1.0, 0.5, 0.0}
\definecolor{amaranth}{rgb}{0.9, 0.17, 0.31}

\graphicspath{{./figures/}}

\def\ltsima{$\; \buildrel < \over \sim \;$}
\def\simlt{\lower.5ex\hbox{\ltsima}}
\def\gtsima{$\; \buildrel > \over \sim \;$}
\def\simgt{\lower.5ex\hbox{\gtsima}}

\newcommand\gwk{\textsc{GWKokab}\xspace}

\newcommand{\svec}{\lambda} 
\newcommand{\comp}{\boldsymbol{\rho}}
\newcommand{\svecz}{\boldsymbol{\lambda}} 
\newcommand{\pvecz}{\boldsymbol{\Lambda}}
\newcommand{\data}{\mathcal{D}}

\def\RIT{Center for Computational Relativity and Gravitation, Rochester Institute of Technology, Rochester, New York 14623, USA}

\begin{document}

\renewcommand{\arraystretch}{1.5}
\title{Uncovering Hierarchical Subpopulations of Binary Black Holes}

\author{M. Zeeshan\orcidlink{0000-0002-6494-7303}}
\email{m.zeeshan5885@gmail.com}
\affiliation{\RIT}

\author{R. O'Shaughnessy\orcidlink{0000-0001-5832-8517}}
\affiliation{\RIT}

\begin{abstract}
The growing gravitational-wave census now contains 259 binary black hole systems
with false-alarm rates below $1~{\rm yr}^{-1}$, enabling studies of trends, structure,
and subpopulations that may reveal their formation mechanisms.
We test whether hierarchical interpretations of previously proposed mass structure
are consistent with the observed spins. We construct a multi-component mixture
model in which each component has independently inferred merger-rate, mass,
spin-magnitude, and spin-orientation distributions, while all components share a
common redshift distribution. The model combines a power law with five Gaussian
mass components whose ordered prior ranges reduce parameter degeneracies.
We recover overall mass and redshift distributions consistent with published
results, as well as the previously identified trend toward low, preferentially
aligned spins at low masses and larger, isotropically oriented spins at high masses.
However, unlike some previous analyses, we find that every Gaussian component
except the lowest-mass component is consistent with isotropic spin orientations.
Conditional on the adopted mass hierarchy, the inferred spins do not follow the
mass--spin correlations expected from hierarchical mergers without additional
spin evolution. Reconciling both distributions may therefore require additional
physics. We present an illustrative, finely tuned spin-down model that could
reconcile a hierarchical interpretation of much of the mass structure with the
observed spin distribution.
\end{abstract}
\maketitle

\section{Introduction}
\label{sec:intro}

Since the first detection of gravitational waves (GW) \cite{2016PhRvL.116f1102A}, the growing census of binary black hole (BBH) mergers \cite{LIGO-O2-Catalog,LIGO-O3-O3b-catalog,LIGO-O4a-cbc-catalog_results,2026arXiv260527090T} identified by ground-based gravitational wave detectors including LIGO,
Virgo, and KAGRA \cite{2015CQGra..32g4001L,2015CQGra..32b4001A,2021PTEP.2021eA101A}
provides increasingly precise information about the population properties of merging compact binaries throughout the universe
\cite{LIGO-O2-Rates,LIGO-O3-O3a-RP,LIGO-O4a-cbc-population,LIGO-GWTC5-populations-2026}.
These observations provide increasingly precise clues to the astrophysical
origins of these binaries  \cite{LIGO-O3-O3bpop,2022LRR....25....1M,2021NatAs...5..749G}, as well as
creating opportunities for GW cosmology through well-characterized mass features that can serve as population-based distance indicators \cite{2022PhRvL.129f1102E,2023PhRvD.108d2002M,2025ApJ...980...85M,2026arXiv260414290G,2026arXiv260315332S,2026A&A...709A.197T}.

The rapidly increasing census has unveiled structure in the joint mass and
spin distribution, suggesting that several physical processes shape the population.
Notably, the distinctive masses, high spin magnitudes, 
and misaligned spin orientations  of high-profile GW observations across the mass spectrum
suggest that a fraction of the GW census
forms hierarchically, through repeated black hole mergers  \cite{LIGO-O4-GW231123,gwastro-agndisk-VeraGW231123McFacts,2025arXiv251113820L,2025arXiv250717551L,LIGO-O4-HierarchicalPair-2025,2025arXiv250923897L,gwastro-mergers-hierarchical-GW231123-Passenger2026,gwastro-mergers-hierarchical-kicks-spins-Borchers2025,gwastro-mergers-hierarchical-redshift-chieff-Farah2026,gwastro-mergers-hierarchical-massratio-chieff-Vijaykumar2026,astrodyn-clusters-bbh-massdistribution-Ye2026,2025arXiv250915646B,gwastro-mergers-AGN-hierarchical-GWTC4-Li2025}.  More broadly, these high-profile events
corroborate analyses performed with the whole census of GW masses and spins
 \cite{LIGO-O4a-cbc-population,2025arXiv250923897L,2025arXiv250717551L,2025arXiv251105316T,2025arXiv251025579T,2025arXiv250915646B,2025PhRvL.134a1401A,2025arXiv250602250S,2024ApJ...960...65S,gwastro-pop-Zeeshan-O4aMixture},
which indicate multiple populations with distinct properties, consistent with a
dynamically and hierarchically formed component predicted theoretically \cite{1995ApJS...99..609S,1998MNRAS.300..857E,MillerHamilton-BHCollisionRunaway2002} and based on analyses with earlier observations
alone \cite{2020ApJ...893...35D,2021ApJ...915L..35K,2025ApJ...981..177L,gwastro-agndisk-GayathriPopModels2025,2022ApJ...927..231F,2025PhRvD.112f3034X,2022PhRvD.106j3013M}.
While a hierarchical origin remains the most compelling explanation, other astrophysical scenarios can also produce some
of these features, including scenarios involving stellar collisions and chemically homogeneous evolution
\citep{astrodyn-clusters-bbh-accretion-massive-spinning-Kiroglu2025,gwastro-mergers-CHE-GW231123-Popa2025}.
No single scenario explains all observations: evidence for dynamical formation across the mass spectrum contrasts with the small spins
inferred for most mergers with well-constrained spins, apart from the most massive systems.

In this work, we test whether a hierarchical interpretation of previously proposed mass structure is consistent
with the observed spins, using the established \gwk inference
framework \cite{gwastro-mergers-zeeshan-gwkokab}  and extending the MultiSource model family widely used to
interpret the GW population \cite{2019phrvd.100d3012w,pop-models-aps-2021,LIGO-O3-O3bpop}. We
specifically adopt a multi-component mixture model with distinct mass and spin distributions, loosely related by hierarchical scales but without requiring each ``generation'' to inherit the remnant
spins (or merger rates) implied by strictly hierarchical formation.  We adopt this mass structure as an ansatz,
rather than evidence for hierarchical formation, and ask whether its independently recovered spins satisfy the
corresponding remnant-spin expectations.  Within this model, we find suppressed hierarchical growth in spin.  Our multi-component model recovers
a preferentially aligned and low-spin distribution for its lowest-mass component; all subsequent populations are consistent with
isotropic spins, with increasing characteristic spin magnitude with mass.  Though we include a power-law component, our model is dominated by the superposition of distinct
Gaussian components associated with naturally hierarchical mass scales.
Our investigation expands upon earlier, preliminary results based on earlier catalogs or a subset of GW observables
that suggested similar hierarchical structure \cite{2022apj...928..155T,2026arXiv260414290G,2026arXiv260525994G,2025arXiv251025579T,2026arXiv260525980Q,2022mnras.509.5454r,2026PhRvD.113h3006T,2026arXiv260600234P,2026arXiv260524281H,2024arXiv240601679P,2025arXiv250915646B,2026arXiv260317987R,2026arXiv260618081T,2021apj...913l..19t,2026arXiv260602318P,2025arXiv250909123A,2025PhRvD.112b3531A,2026arXiv260407456G}.

The paper is organized as follows.
In Section~\ref{sec:methods}, we review our framework for hierarchical population
inference; introduce our population model; and describe the GW observations and surveys incorporated within our
inference.
In Section~\ref{sec:results}, we describe the immediate results of fitting our multi-component model to the GW census:  the recovered mass distribution both overall (matching previously published
results) and per-component, unveiling hierarchical structure; and the recovered spin distribution of each component,
contrasting our results with previous claims about spin orientations versus mass.
In Section~\ref{sec:discuss}, we reflect on the astrophysical interpretation of our model, using quantities derived from our multi-component model.
We conclude in Section~\ref{sec:conclude}.

\section{Methods}
\label{sec:methods}
\subsection{Review of hierarchical Bayesian inference}

To infer the BBH population, we use the Bayesian parametric modeling formalism
implemented in \gwk \cite{gwkokab2024github,2026PhRvD.113j3003Q}.
Given the likelihoods $\ell_j(\svecz)$ for individual sources and their reference
priors $\pi(\svecz)$, we infer the population hyperparameters $\pvecz$ within
the hierarchical Bayesian framework of \Cref{eq:Bayes_pop}:

\begin{align}
    \label{eq:Bayes_pop}
    \!p\left(\pvecz | \data \right)
     & = \frac{
        \pi(\pvecz)\,
        p(\data | \pvecz)
    }{
        p\left( \data \right)
    },
\end{align}
where $\data=\{d_j\}_{j=1}^N$ denotes the data for $N$ events, and
$\pi(\pvecz)$ is the prior on the population hyperparameters.
The evidence $p(\data)$ normalizes the posterior and can be omitted when sampling
$\pvecz$. Defining $\mathcal{L}(\pvecz)\equiv p(\data\mid\pvecz)$, we therefore
sample $p(\pvecz\mid\data)\propto\mathcal{L}(\pvecz)\pi(\pvecz)$.

We model the detections as an inhomogeneous Poisson process \cite{2019MNRAS.486.1086M,2004AIPC..735..195L,PhysRevD.91.023005}
\begin{equation}
    \label{eq:likelihood}
    \mathcal{L}(\pvecz) \propto
    e^{-\mu{(\pvecz)}}
    \prod_{j=1}^N
    \int\ell_j(\svecz) \cdot \comp(\svecz\mid\pvecz)
     \sqrt{ g_{\svecz}}
    d \svecz,
\end{equation}
where $\mu(\pvecz)$ is the expected number of detections for population
hyperparameters $\pvecz$, as defined in \Cref{eq:mu}.
The quantity $g_{\svecz}$ is the determinant of the metric on parameter space,
and $\comp(\svecz\mid\pvecz)$ is the source-frame merger-rate density.
The integration measure contains the observing time $T_{\mathrm{obs}}$, the
comoving-volume element $dV_c/dz$, the source-to-detector time conversion,
and the mass and spin coordinate measures:
$\sqrt{g_{\svecz}}d\svecz = T_{\mathrm{obs}}\,dz\,(1+z)^{-1}(dV_c/dz)\,dm_1\,dm_2$
multiplied by the appropriate spin-coordinate factors.
The event likelihood is related to its posterior and reference prior by

\begin{equation}
\ell_j(\svecz) \equiv p(d_j|\svecz) \propto \frac{ p(\svecz|d_j)}{\pi(\svecz)}.\label{eq:indi_likelihood}
\end{equation}
The reference prior $\pi(\svecz)$ is taken from the corresponding parameter-estimation (PE) analysis. For analyses using a prior uniform in comoving volume and source-frame time, the mass--redshift contribution is
\begin{equation}
    \pi(\svecz)=\frac{1}{V_0}\frac{dV_c}{dz}\frac{1}{1+z}\times(1+z)^2,
    \label{eq:ref_prior}
\end{equation}
where $(1+z)^2$ is the Jacobian for converting both component masses from the detector frame to the source frame, and $V_0$ is the normalization of the comoving-volume prior. Further details are given
in \cite{2021arXiv210409508C}. All integrals appearing explicitly or implicitly in the expressions are computed via Monte Carlo integration, as described in \cite{2026PhRvD.113j3003Q}. We retain hyperparameter samples only when the estimated likelihood variance is less than 1; see Eqs.~(9)--(11) of \cite{2025PhRvD.111f3043H}.

The expected number of GW detections can be formulated as an integral over the intrinsic source-parameter space $\svec$ and redshift $z$ modulated by an appropriate selection (weighting) function. The total expected number of detections summing over all populations is given by

\begin{align}
    \label{eq:mu}
    \mu(\pvecz)= \int P_{\mathrm{det}} (\svec;z)\cdot \comp(\svecz\mid\pvecz)\sqrt{ g_{\svecz}}
    d \svecz.
\end{align}
Here $P_{\mathrm{det}}(\svec;z)$ is the detection probability for a source with intrinsic parameters $\svec$ at redshift $z$.

\subsection{Population Model and Priors}
\label{subsec:pop_model}

In this study, we construct a multi-component population model to capture the hierarchical subpopulation of binary
black holes. The model consists of a smoothed power-law (SPL) component describing low-spin binaries
and five Gaussian mass components describing distinct subpopulations. We fix the maximum mass $m_{\max}$ to
$300\,M_\odot$ for all components of the model.   We infer minimum primary and secondary masses shared by all components.
Each component of the model has its own independent spin and tilt distribution, but the redshift distribution is shared among all components \cite{2026arXiv260527226T}.
We use Gaussian spin-magnitude distributions truncated to $[0,1]$. For the two spin tilts, expressed as
$\cos(\theta)$, we adopt a mixture of a Gaussian truncated to $[-1,1]$
and an isotropic distribution. The two tilts are identically distributed but are not assumed to be independent. For the redshift distribution, we adopt a power-law model with a single parameter $\kappa$ to capture the redshift evolution of the merger rates. The mathematical formulation of the model is given in \Cref{eq:pop_model}.
\begin{align}
    \label{eq:pop_model}
    \comp(\svecz\mid\pvecz) & =  (1+z)^\kappa \cdot \sum_{k=0}^5 \comp_k(\svecz\mid\pvecz)
\end{align}
where $\comp_k(\svecz\mid\pvecz)$ is the merger rate density of the $k$th component of the model. The first component of the model is an SPL and the remaining five components are truncated Gaussian distributions. The SPL component is defined by
\begin{align}
  \label{eq:smooth_pl}
    \comp_0(\svecz\mid\pvecz) &=   m_1^{-\alpha_{\mathrm{spl}}} q^{\beta_{\mathrm{spl}}} \pi(\chi) \pi(\cos\theta) \nonumber\\ &\times S(m_1\mid \delta_{m_1,{\rm spl}}, m_{1,\min}) \times \mathcal{R}_{\mathrm{spl}}.
\end{align}
Expressions for the smoothing function $S(m_1\mid \delta_{m_1,{\rm spl}}, m_{1,\min})$ and the spin distributions $\pi(\chi)$ and $\pi(\cos\theta)$ are given in Eqs.~(B13), (B15), and (B16), respectively, of the GWTC-5 population paper \cite{LIGO-O4b-CBC-pop}.
For each Gaussian mass component from $k=1$ to $k=5$, we model the primary and secondary source-frame masses using independent one-dimensional truncated normal distributions,

\begin{align}
\comp_k(m_1,m_2)
&=
\mathcal{N}_{\rm k}
\!\left(
 m_1 \mid
 \mu_{m,{\rm gg},k},
 \sigma_{m,{\rm gg},k},
 m_{1,k}^{\rm low},
 m_{1,k}^{\rm high}
\right)
\nonumber\\
&\times
\mathcal{N}_{\rm k}
\!\left(
 m_2 \mid
 \mu_{m,{\rm gg},k},
 \sigma_{m,{\rm gg},k},
 m_{2,k}^{\rm low},
 m_{2,k}^{\rm high}
\right)
\nonumber\\
&\times
\pi_k(\chi)\,\pi_k(\cos\theta) \times \mathcal{R}_{G_k}.
\label{eq:gaussian_component_mass}
\end{align}

Thus, the Gaussian mass components are separable in $m_1$ and $m_2$;
they are not modeled as a correlated two-dimensional Gaussian in the ($m_1,m_2$) plane. The prior ranges for all parameters of the multi-component population model are listed in \Cref{tab:population_priors}.

\begin{table*}
\caption{
Prior ranges for the multi-component population model. Here
$\mathcal{U}(a,b)$ denotes a uniform prior on the interval $[a,b]$.
Parameters for the secondary spin magnitude and tilt distributions are tied
to the corresponding primary-spin parameters unless listed separately.
}
\label{tab:population_priors}
\begin{ruledtabular}
\begin{tabular}{llll}
Model Components & Parameter & Prior & Description \\
\hline

Smoothed power law
& $\alpha_{\rm spl}$
& $\mathcal{U}(-4,12)$
& Primary-mass power-law index \\

& $\beta_{\rm spl}$
& $\mathcal{U}(-2,7)$
& Mass-ratio power-law index \\

& $\delta_{m_1,{\rm spl}},\delta_{m_2,{\rm spl}}$
& $\mathcal{U}(0,10)$
& Low-mass smoothing scales \\

& $m_{1,\min}$
& $\mathcal{U}(2,10)\,M_\odot$
& Minimum primary mass \\

& $m_{2,\min}$
& $\mathcal{U}(2,m_{1,\min})\,M_\odot$
& Minimum secondary mass \\ \hline

Low-spin SPL
& $\mu_{\chi,{\rm spl}}$
& $\mathcal{U}(0,0.15)$
& Spin-magnitude location \\

& $\sigma_{\chi,{\rm spl}}$
& $\mathcal{U}(0.005,0.15)$
& Spin-magnitude width \\ \hline

Gaussian masses
& $\mu_{m,{\rm gg},1}$
& $\mathcal{U}(5,13)\,M_\odot$
& Primary and secondary location of $G_1$ \\

& $\mu_{m,{\rm gg},2}$
& $\mathcal{U}(13,25)\,M_\odot$
& Primary and secondary location of $G_2$ \\

& $\mu_{m,{\rm gg},3}$
& $\mathcal{U}(25,45)\,M_\odot$
& Primary and secondary location of $G_3$ \\

& $\mu_{m,{\rm gg},4}$
& $\mathcal{U}(45,65)\,M_\odot$
& Primary and secondary location of $G_4$ \\

& $\mu_{m,{\rm gg},5}$
& $\mathcal{U}(65,90)\,M_\odot$
& Primary and secondary location of $G_5$ \\

& $\sigma_{m,{\rm gg},k}$
& $\mathcal{U}(0,10)\,M_\odot$
& Mass width for Gaussian components $k=1,2,3$ \\

& $\sigma_{m,{\rm gg},k}$
& $\mathcal{U}(0,15)\,M_\odot$
& Mass width for all Gaussian components $k=4,5$ \\
\hline

Spin magnitudes
& $\mu_{\chi,{\rm gg},k}$
& $\mathcal{U}(0,1)$
& Spin-magnitude location for $k=1,\ldots,5$ \\

& $\sigma_{\chi,{\rm gg},k}$
& $\mathcal{U}(0.005,1)$
& Spin-magnitude width for $k=1,\ldots,5$ \\ \hline

Spin tilts
& $\zeta_{\cos\theta,{\rm (spl,gg)},k}$
& $\mathcal{U}(0,1)$
& Tilt-mixture fraction for $k=0,\ldots,5$ \\

& $\mu_{\cos\theta,{\rm (spl,gg)},k}$
& $\mathcal{U}(-1,1)$
& Tilt-location parameter for $k=0,\ldots,5$ \\

& $\sigma_{\cos\theta,{\rm (spl,gg)},k}$
& $\mathcal{U}(0.01,4)$
& Tilt-width parameter for $k=0,\ldots,5$ \\ \hline

Log Merger rate
& $\ln \mathcal{R}_{G_k}$, $\ln \mathcal{R}_{spl}$
& $\mathcal{U}(-10,10)$
& Power-law and Gaussian components $k=0,\ldots,5$ \\

Redshift evolution
& $\kappa$
& $\mathcal{U}(-10,10)$
& Redshift-evolution parameter of the full model\\

\end{tabular}
\end{ruledtabular}
\end{table*}

\subsection{GW Surveys and Observations}
We include all BBH events identified through GWTC-5 that satisfy the selection criterion
$\mathrm{FAR}<1~\mathrm{yr}^{-1}$, using the minimum false-alarm rate across search pipelines, and characterized in either GWTC-2 \cite{LIGO-O3-O3a-catalog}, GWTC-3 \cite{LIGO-O3-O3b-catalog},
GWTC-2.1 \cite{LIGO-O3-O3a_final-catalog},  GWTC-4 \cite{LIGO-O4a-cbc-catalog_results}, or GWTC-5 \cite{LIGO-O4b-CBC-catalog},
resulting in
259 BBH events from O1 through O4b.
We use published LVK posterior samples and apply weights as needed to account for
differences in PE priors across observing runs. For each event, we select 5000
posterior samples for population inference, using the MIXED data product whenever it is available.

To account for survey selection effects, we use previously reported search sensitivity
estimates \cite{2025arXiv250818081T,2025PhRvD.112j2001E}, selecting injections with SNR above 10 \cite{2026arXiv260527226T}.\footnote{For GWTC-5, we use the semi-analytical sensitivity
injections published as \url{mixture-semi_o1_o2-real_o3_o4a_o4b-cartesian_spins_20260327234151UTC.hdf}.}

\section{Results}
\label{sec:results}

\subsection{Mass distribution}
\begin{figure*}[ht]
\centering
\includegraphics[width=0.50\linewidth]{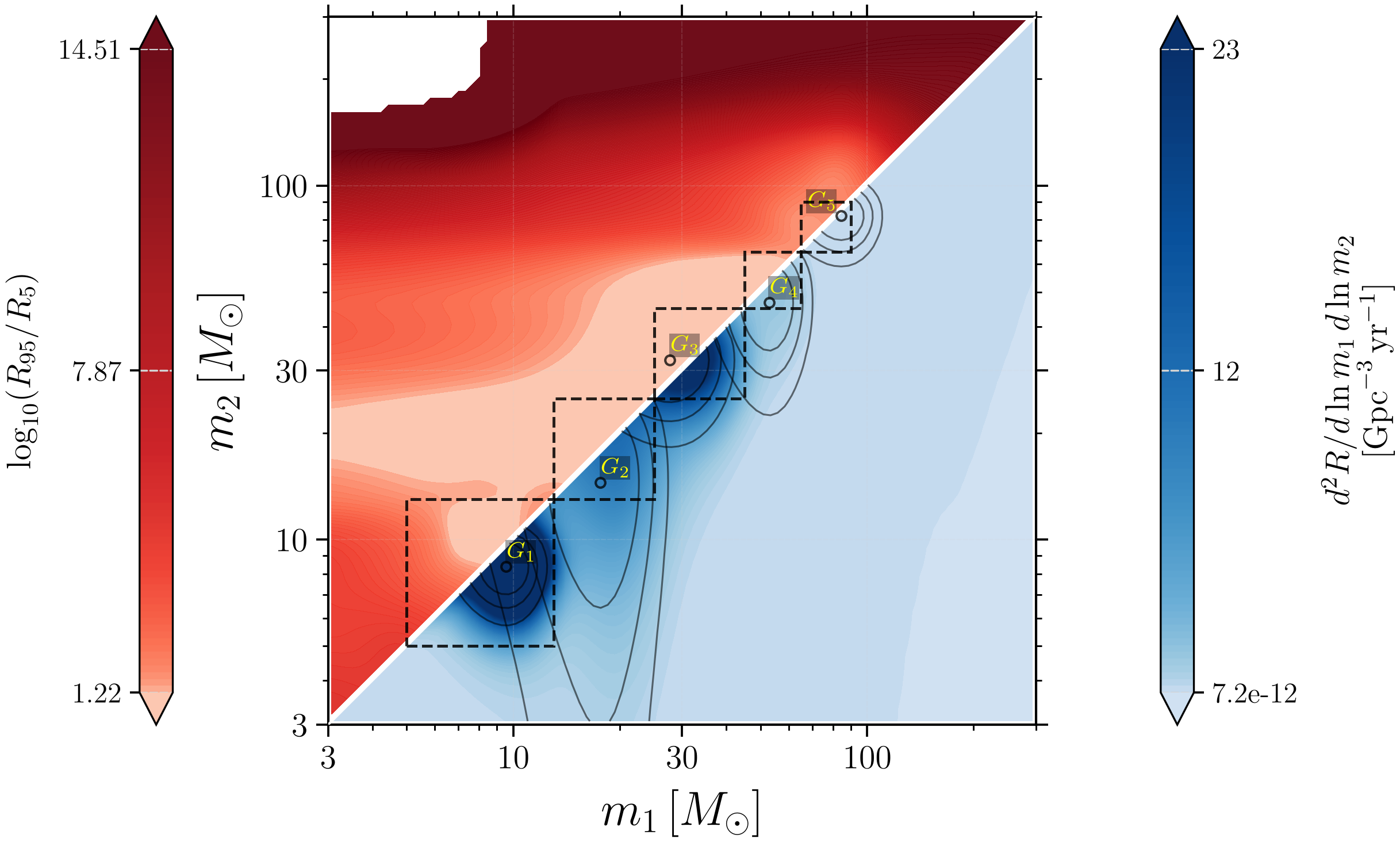}
\includegraphics[width=0.44\linewidth]{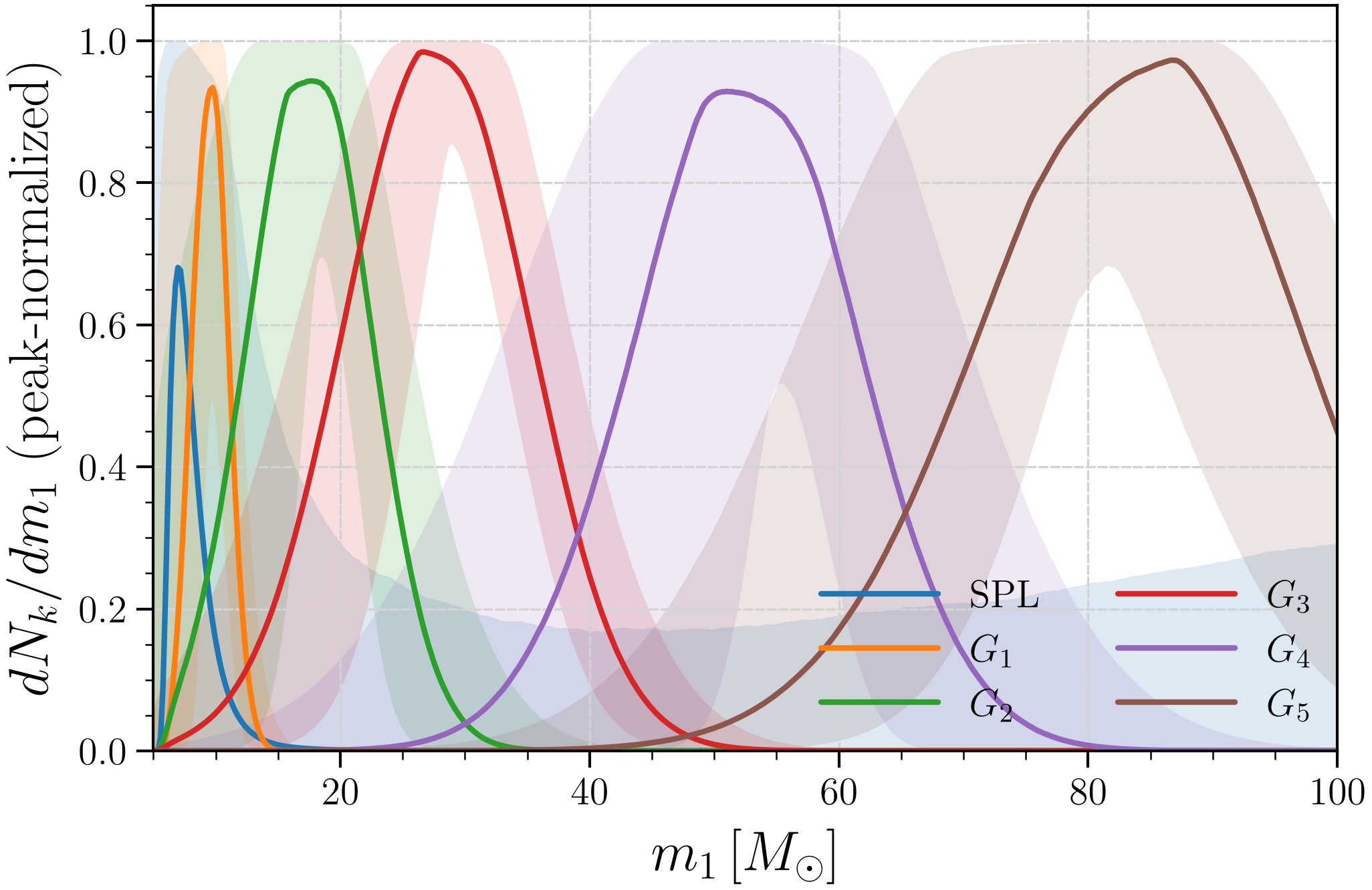}
\caption{The left panel shows the inferred merger-rate density in the $(m_1,m_2)$ plane at $z=0$. The lower triangle displays the posterior median of $d^2R/(d\ln m_1,d\ln m_2)$, while the upper triangle shows the logarithmic uncertainty, $\log_{10}(R_{95}/R_5)$, where $R_5$ and $R_{95}$ denote the 5th and 95th posterior percentiles, respectively. Regions with negligible median rate density are masked. Black contours indicate the posterior-median locations and shapes of the Gaussian mass components $G_1,\ldots,G_5$ shown individually. The displayed maps are smoothed for visualization, and dashed lines mark the prior ranges of the Gaussian component centers. The right panel shows peak-normalized primary-mass distributions for the inferred subpopulations, highlighting differences in characteristic mass scales and widths while removing information about their relative merger rates.}
\label{fig:density_results}
\end{figure*}

Figure \ref{fig:density_results} shows the median of the merger rate density versus binary component
masses.  The labels and contours indicate the contributions from each of the five Gaussian components which dominate the
overall population.   The upper triangle shows the uncertainty in the merger-rate density. It is small where many events are
observed (near $10M_\odot$ for $G_1$ and near $30M_\odot$ for $G_3$) and
large in regions with few observations (e.g., highly asymmetric binaries with
$m_1\simeq 100M_\odot$).

Our multi-component model imposes informative prior ranges on the location of each Gaussian component, motivated by
previous observations. These constraints also reduce the degeneracies inherent in flexible mixture models.
Partly by construction, the Gaussian components cluster near the equal-mass line $m_1=m_2$ in both peak location and extent,
except for $G_2$, which has a larger relative extent in secondary mass. The inferred mean and width of each
Gaussian are localized within their prior ranges.

Integrating over our population, we find an overall BBH merger rate at $z=0$ in the range $(19.6-59.2){\rm Gpc}^{-3}{\rm
yr}^{-1}$, and at $z=0.2$ in the range $(31.4-90.2){\rm Gpc}^{-3}{\rm
yr}^{-1}$, slightly higher than the GWTC-5 estimate \cite{2026arXiv260527226T}. The difference may reflect the greater flexibility of our model.

\begin{figure*}[ht]
\centering
\includegraphics[width=0.49\linewidth]{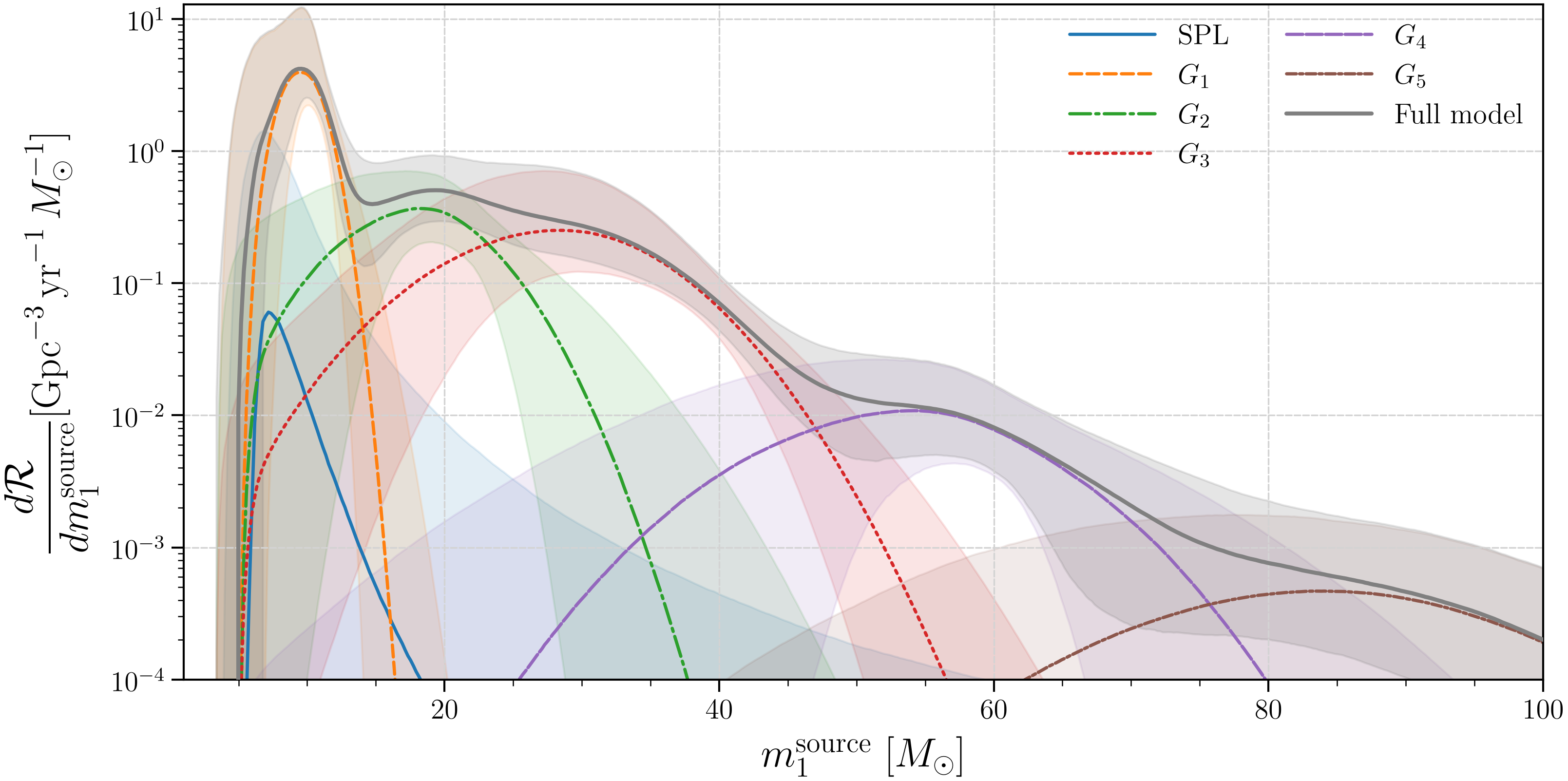}
\includegraphics[width=0.48\linewidth]{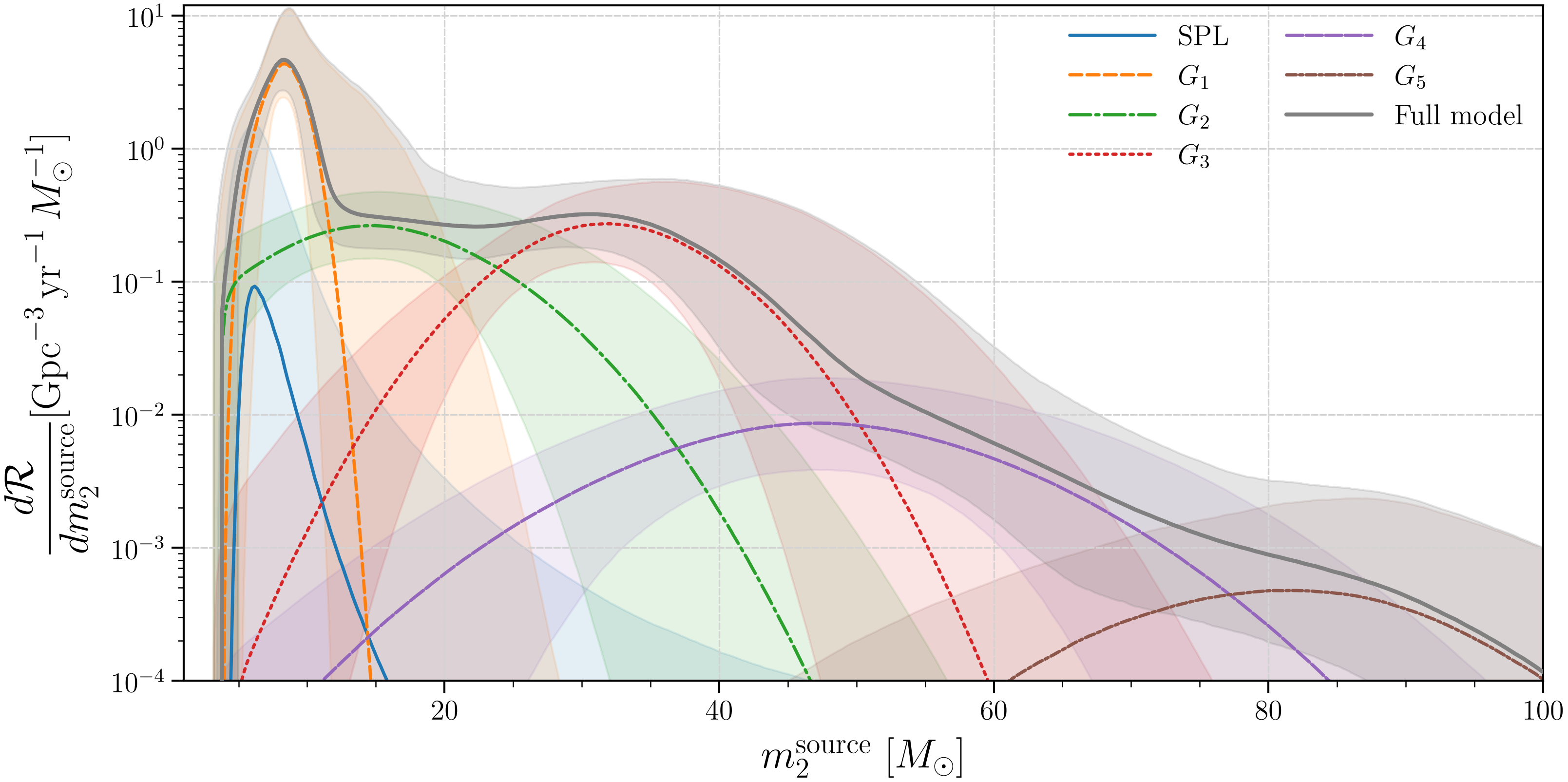}
\caption{The left (right) figure shows the marginal distributions of primary (secondary) mass. The solid gray line represents the posterior median for the full model, while the shaded regions indicate the 90\% credible intervals. The colored lines represent the contribution of each component of the model to the overall mass distribution. The low-spin SPL component is shown in blue, while the Gaussian components are shown in different colors. The figure illustrates how each component contributes to the overall mass distribution of binary black holes in the GWTC-5 catalog.}
\label{fig:mass_results_component}
\end{figure*}

\begin{figure*}[ht]
\centering
\includegraphics[width=0.49\linewidth]{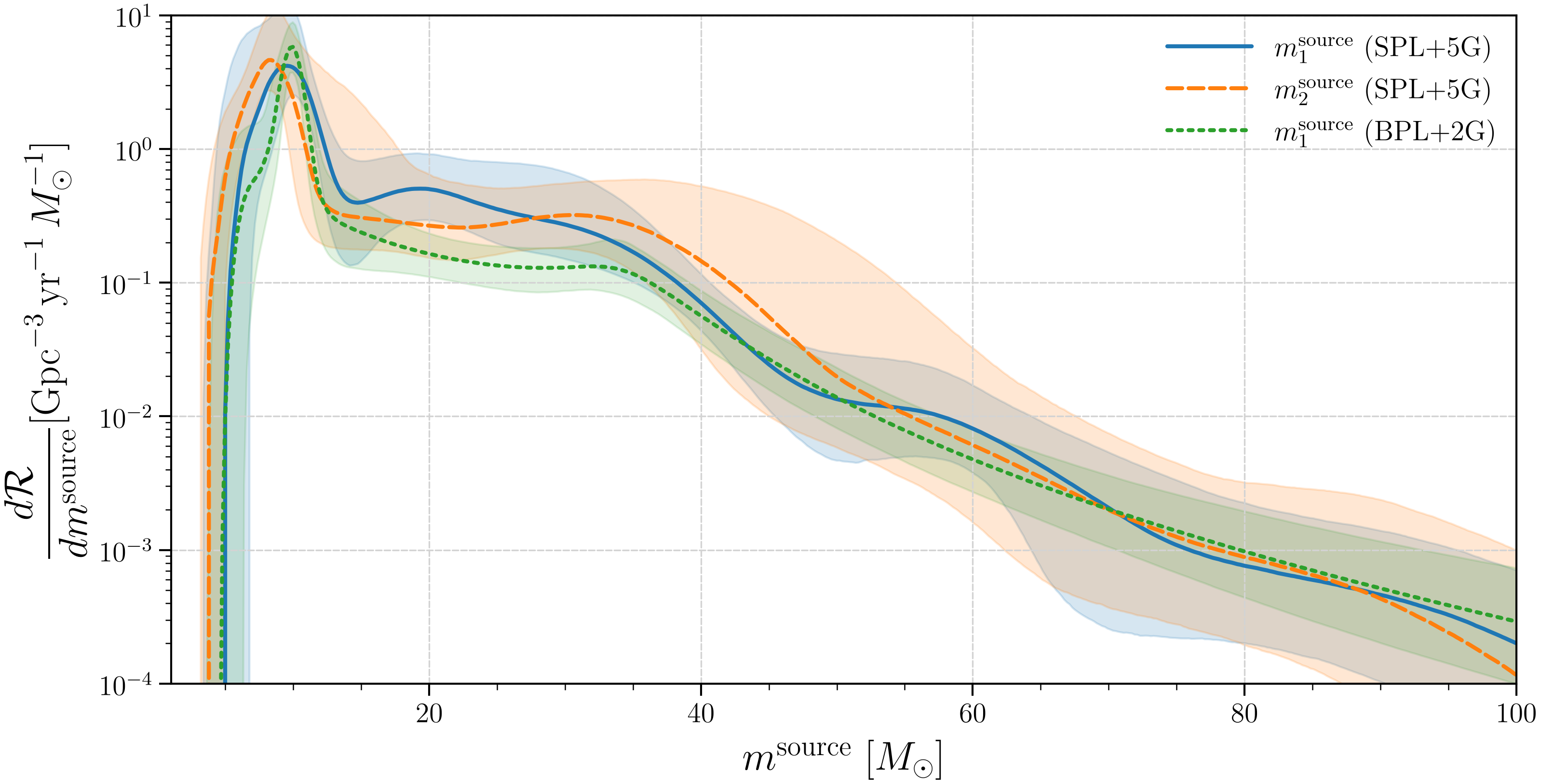}
\includegraphics[width=0.48\linewidth]{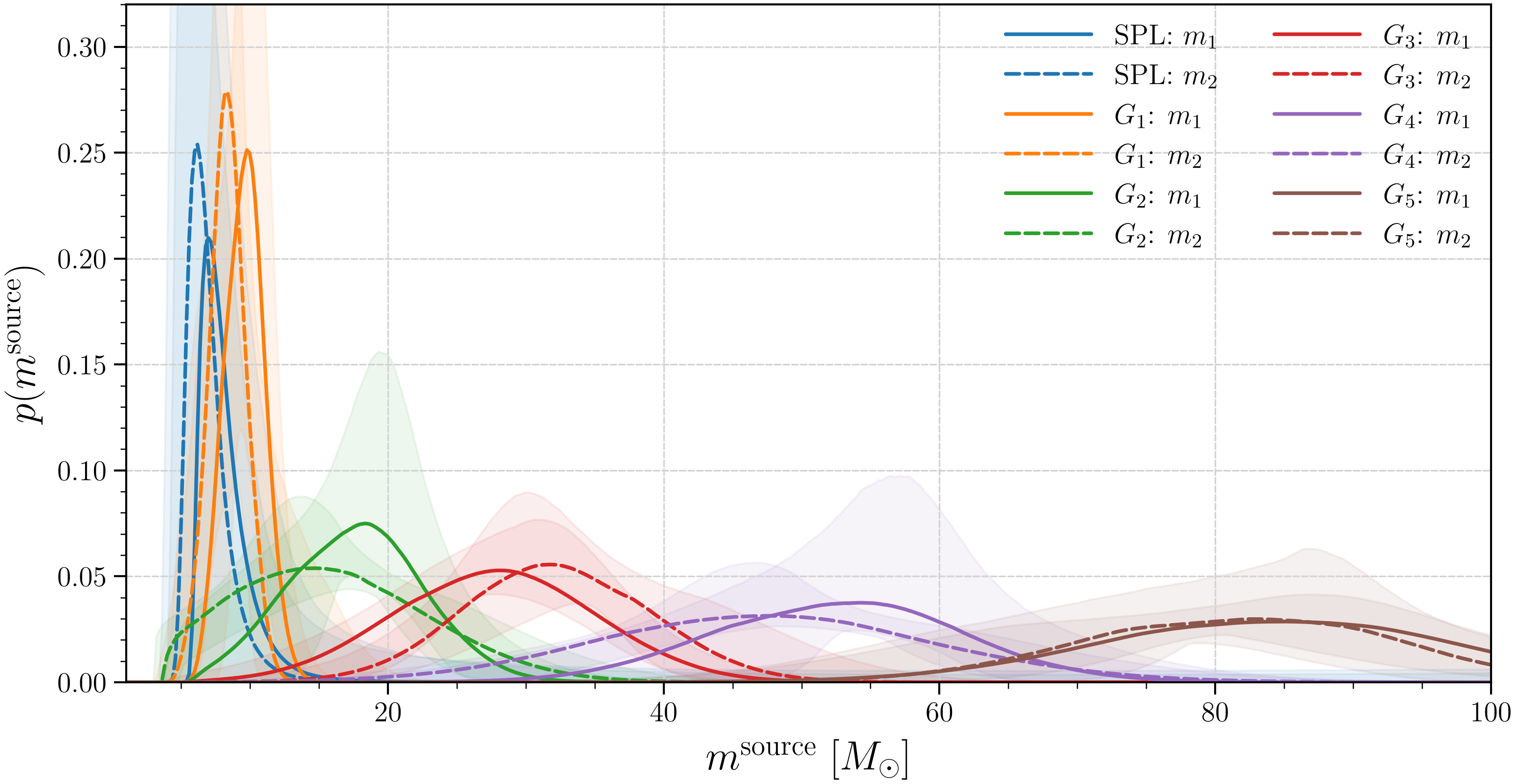}
\caption{The left figure shows the posterior distributions of the full model for primary and secondary masses. The solid (dashed) lines represent the median of the posterior distribution of primary (secondary) masses, and the green dotted curves show the default O4b BBH model (a broken power law plus two Gaussians). The right figure shows the component shapes of the multi-component population model overlaid on the posterior distributions of primary and secondary masses. The shaded regions show the 90\% credible intervals.}
\label{fig:mass_results}
\end{figure*}

Figure \ref{fig:mass_results_component} shows the merger rate versus primary and secondary mass (in gray). This figure
also shows the contributions from each component in our mixture model, illustrating how each component dominates in a
selected mass region.
The power-law component contributes negligibly except at the lowest masses described by our model.  Additionally, the
primary and secondary masses are well-described with similar distributions, both overall and for each component.
Our model does not identify  ``gaps'' where the merger rate is negligible along the line $m_1=m_2$;
cf. \cite{2017ApJ...851L..25F,2021ApJ...909L..23E,2021ApJ...913L..23E,2022ApJ...931..108F,2026ApJ...998L..20R,2026Natur.652..874T,2026NatAs.tmp..111A}.
Figure \ref{fig:mass_results} represents the same information, rearranged to simplify comparison between the
distributions of primary and secondary mass.
This figure demonstrates that with few exceptions, the primary- and secondary-mass distributions are similar, both overall and within \emph{each individual component}.  That said, the Gaussian components $G_2$, $G_3$, and particularly $G_4$ suggest some consistent asymmetry between the
merging binaries, peaking slightly away from equal mass.  Other investigations have previously suggested the compact
binary population could preferentially be asymmetric  for some or all
binaries \cite{2023arxiv230401288G,2024ApJ...962...69F}.
Consistent with GWTC-5, we do not see significant evidence for asymmetry in the $10M_\odot$ peak.
However, in qualitative agreement with previous work \cite{2024MNRAS.527..298T,2026arXiv260614472F}, we do see clear trends
in the range of asymmetry allowed versus mass.  That said, in the absence of detections in the high-mass-ratio regime,
by construction our model does not incorporate unconstrained Gaussian components filling the high-mass-ratio space; thus
we do not assess trends versus mass ratio far from the equal-mass line, in contrast to other work whose models extend to
this region and provide upper limits.

Strikingly in Figure \ref{fig:mass_results}, all of our Gaussian components are comparably broad (except for the lowest-mass BHs in the $10M_\odot$ peak $G_1$).

\subsection{Spin distribution}
\begin{figure*}[ht]
\centering
\includegraphics[width=0.49\linewidth]{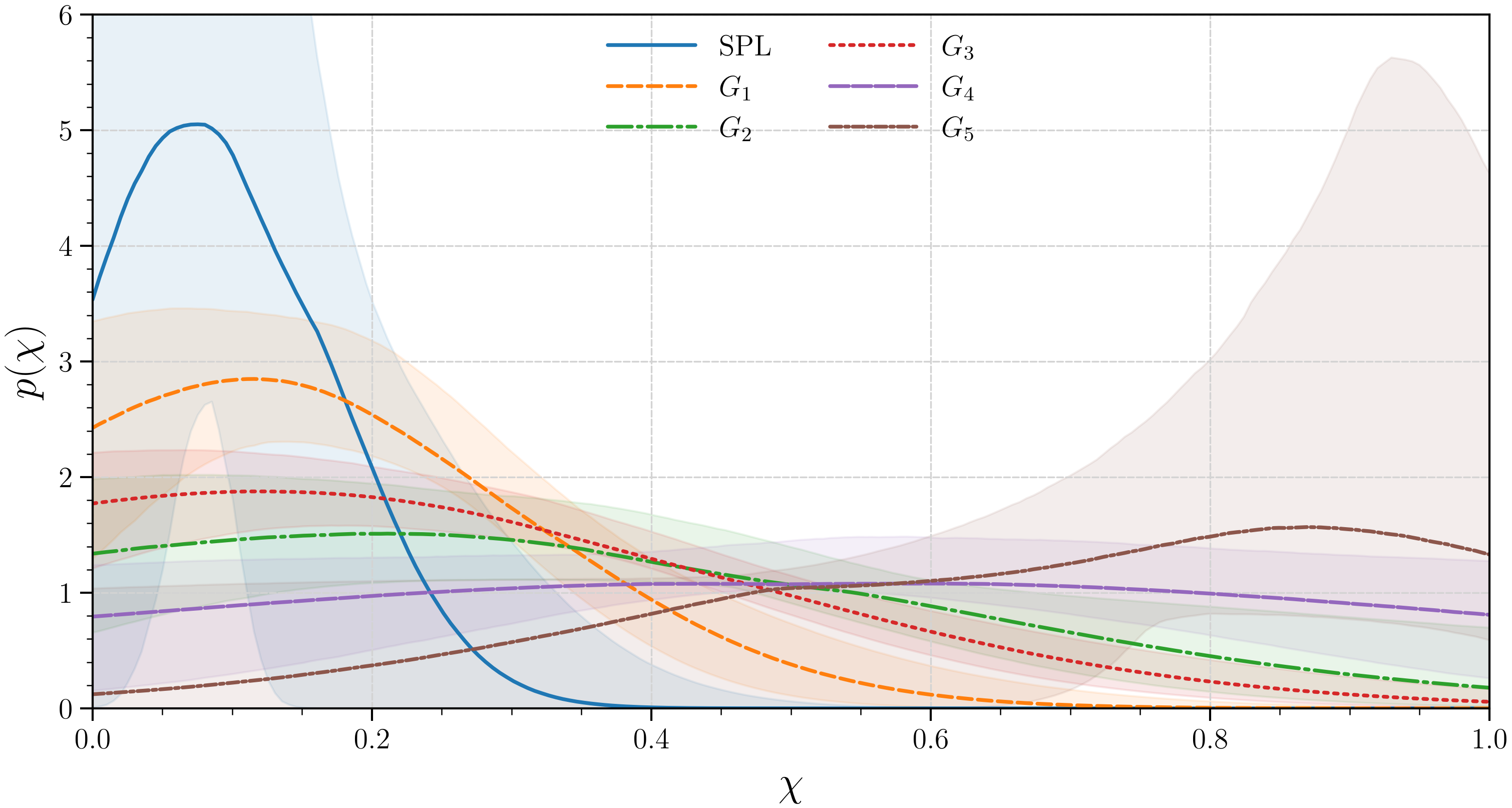}
\includegraphics[width=0.49\linewidth]{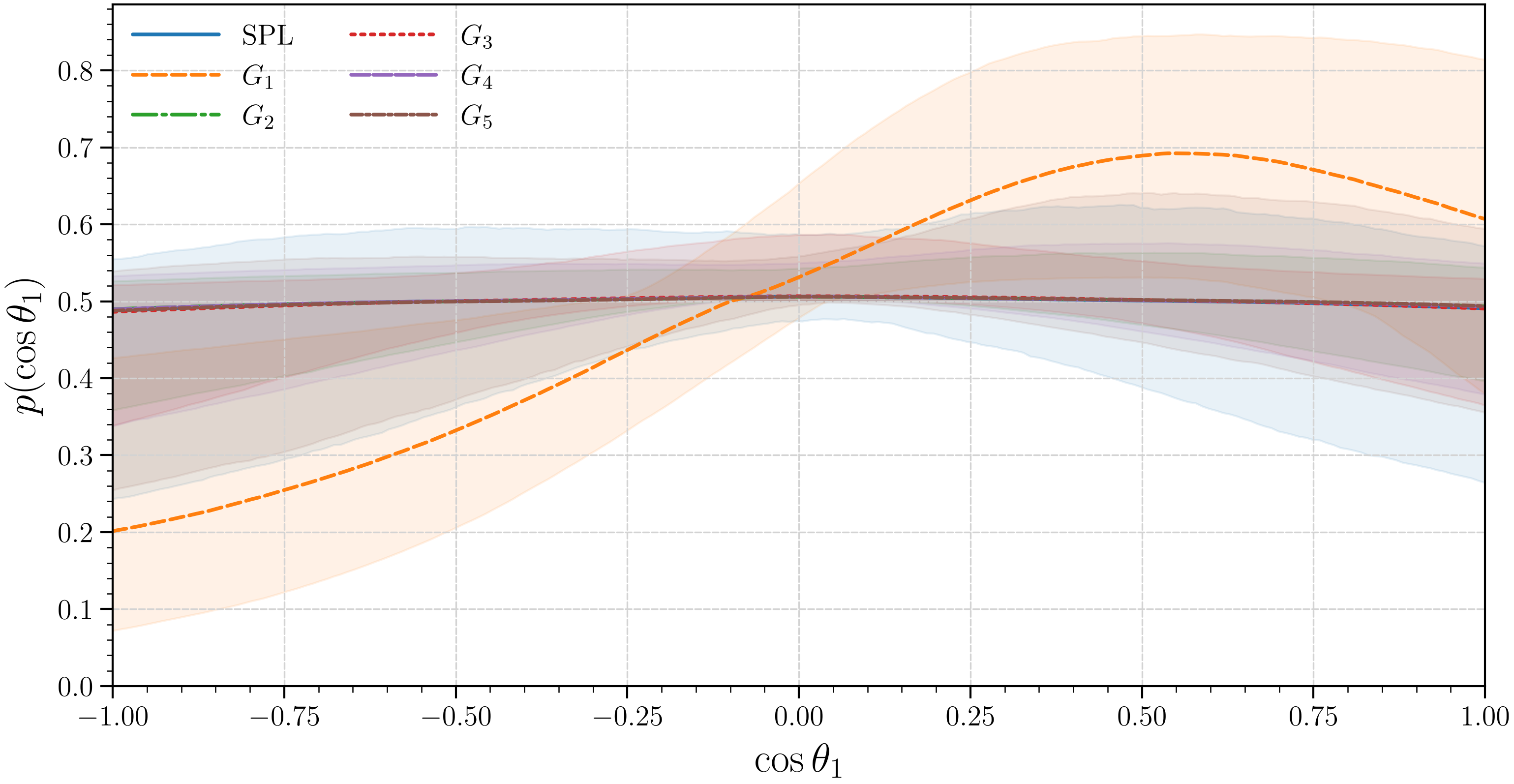}
\caption{The left figure shows the posterior distributions of the spin magnitudes of each component. The curves show posterior medians, and the shaded regions indicate 90\% credible intervals. The right panel shows the spin-tilt distribution of each component.}
\label{fig:spin_results}
\end{figure*}

Our multi-component model naturally unveils trends in spin magnitude and alignment versus mass.
Figure   \ref{fig:spin_results} shows the spin magnitude and alignment distribution inferred for each component.
Consistent with past investigations \cite{LIGO-GWTC5-populations-2026}, we find that low-mass binaries in the power-law component and
the lowest-mass Gaussian have modest spins.  For the lowest-mass Gaussian $G_1$, our inferences strongly favor an
aligned population.  As discussed extensively in prior work, these two signatures can arise naturally from isolated
binary evolution  \cite{2010CQGra..27k4007M,2017MNRAS.471.2801S,gwastro-ConstrainChannels-BoxingDayKicks-Me2017,gwastro-DanielW-PopsynKickPaper2017,2020ApJ...895..128M,2021ApJ...921L..15G,2024ApJ...966L..16P,2023ApJ...946...50B,2026PhRvD.113d3048B}.
The remaining Gaussian components are consistent with isotropic spin orientations and generally have larger characteristic spins at higher masses, although the spins are substantially lower than expected for
merger remnants of lower-mass progenitors without subsequent spin evolution. The most massive Gaussian component
favors large spin magnitudes.
Apart from the unexpectedly low spin magnitudes, these signatures are
qualitatively compatible with hierarchical compact binary formation.

Recently, several studies have characterized trends of spin alignment versus mass with an  ``isotropic fraction''
\cite{2025arXiv251105316T,2026arXiv260107908P,LIGO-GWTC5-populations-2026}. These analyses disfavor spin-isotropic
mergers near $10M_\odot$, consistent with our $G_1$ population which is preferentially low-spin and aligned. 
These calculations also favor a non-isotropic component between $25-40 M_\odot$; by contrast, within the context of our
calculations, we do not see evidence favoring aligned mergers in this mass region ($G_3$), notably including the
frequently detected binaries with masses  $\simeq 30M_\odot$.  The discrepancy between these results could arise from
different assumptions: these ``isotropic fraction'' calculations make extremely strong assumptions about the spin
distribution required for hierarchical mergers, which do not resemble the outcomes seen in the $G_2$ peak. As a result,
by construction these models'  strong assumptions about spin \emph{magnitude} may drive their interpretation. By
contrast, our calculations independently allow both spin magnitude and orientation to vary freely in each component.  
A firm resolution of this discrepancy requires more observations, such that our more flexible model is
better constrained by available data.

\begin{figure*}[ht]
\centering
\includegraphics[width=0.49\linewidth]{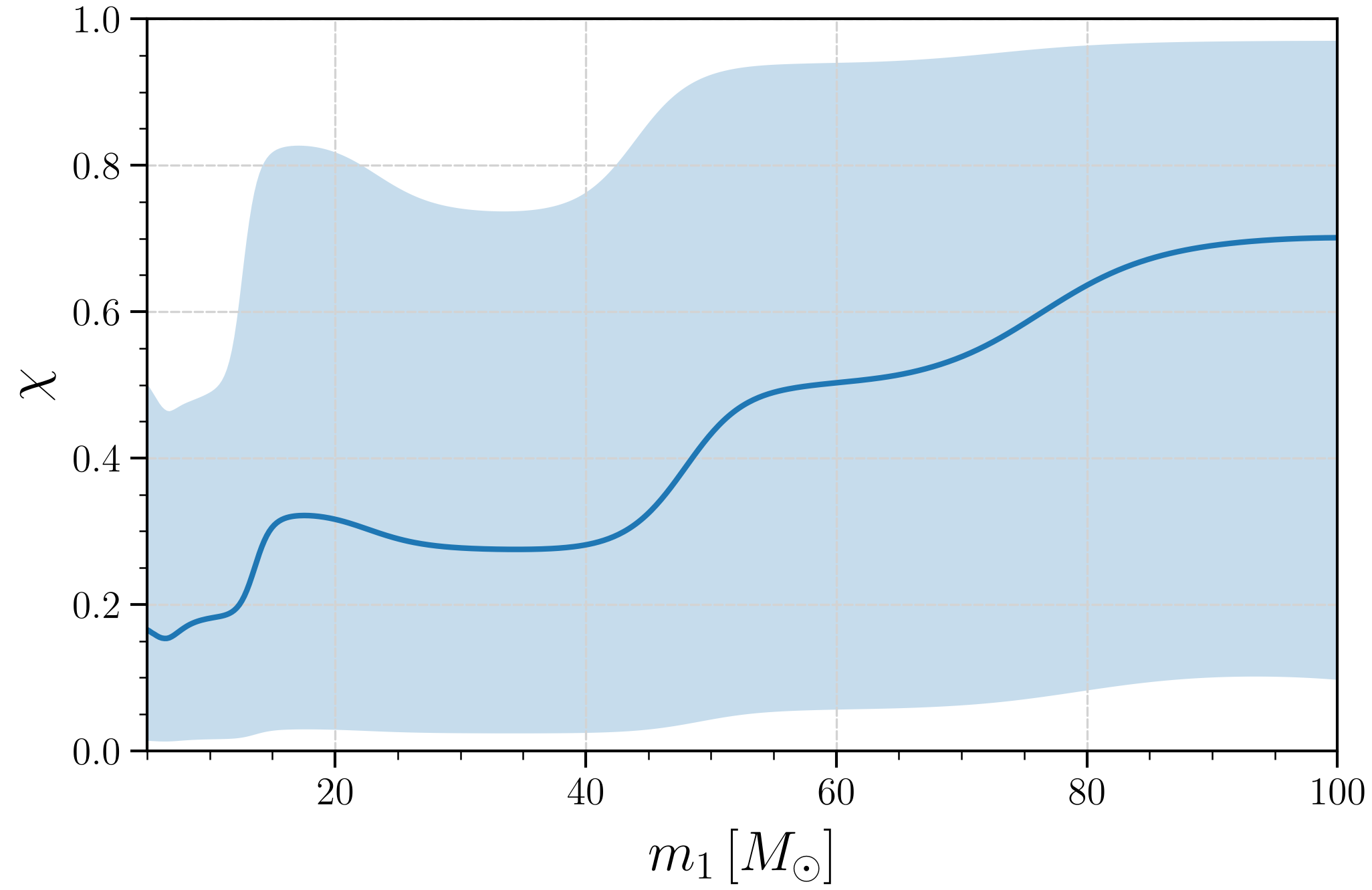}
\includegraphics[width=0.49\linewidth]{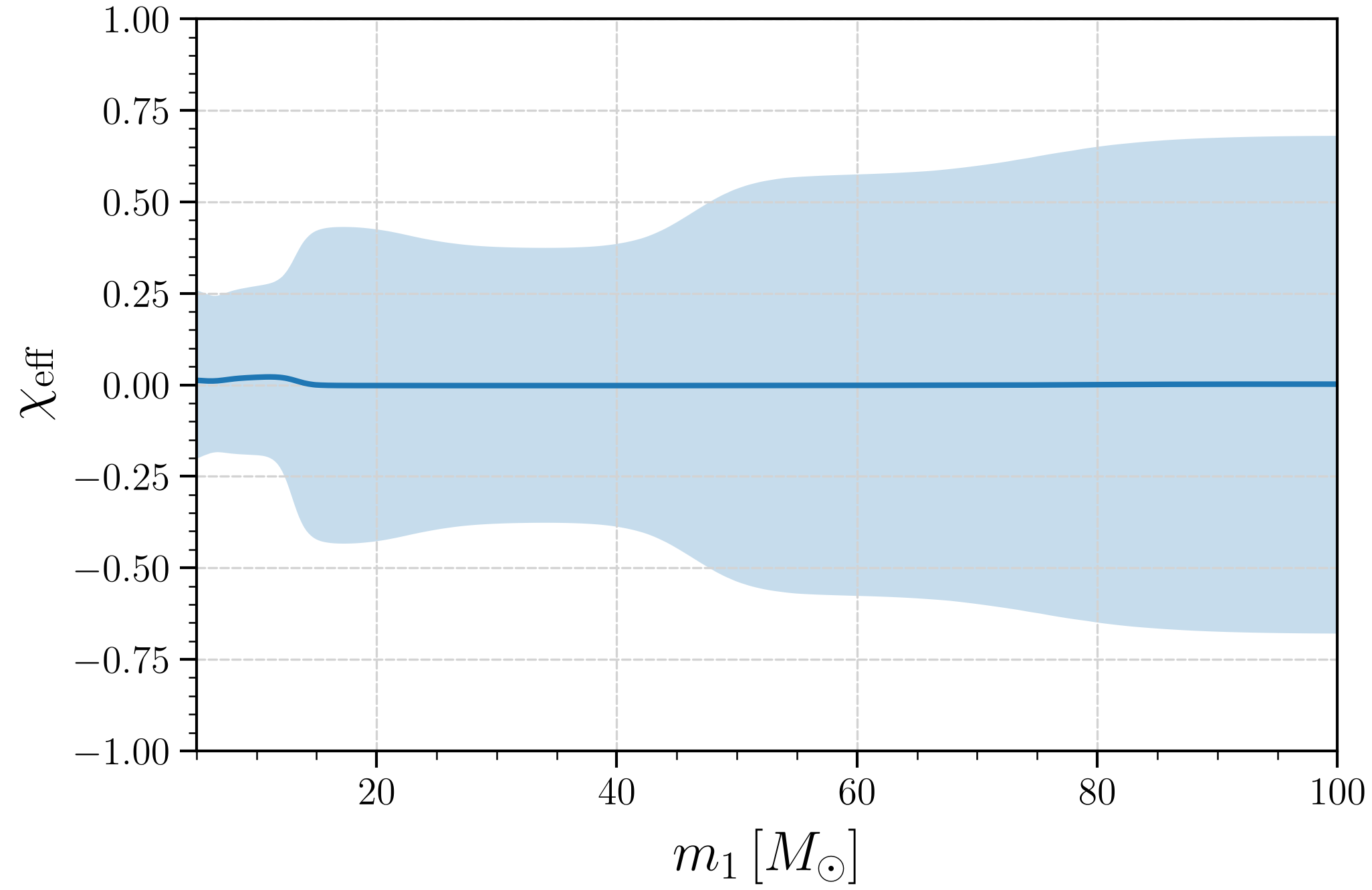}
\caption{The left figure shows the distribution of the spin magnitude conditioned on primary mass, $p(\chi \mid m_1)$. The solid curve shows the median spin magnitude inferred from the posterior-mean joint distribution $p(m_1,\chi)$, while the shaded region shows the central 90\% interval. The right panel shows the posterior predictive distribution of effective spin conditioned on primary mass, $p(\chi_{\rm eff}\mid m_1)$. For each posterior sample, we compute the 5\%, 50\%, and 95\% conditional quantiles of $\chi_{\rm eff}$ at fixed $m_1$, then summarize these quantile curves across posterior samples.}
\label{fig:mas_vs_spin_results}
\end{figure*}

Figure~\ref{fig:mas_vs_spin_results} shows how the spin distributions vary with primary mass.
The left panel shows the conditional spin-magnitude distribution. Its median increases with mass,
consistent with previous studies \cite{2020ApJ...894..129S,2025arXiv251025579T,2025arXiv251105316T,2025arXiv250915646B,2026PhRvD.113d3048B}.
The right panel shows the conditional effective-spin distribution, which broadens with mass,
as also found in previous work \cite{2022apj...928..155T,2025PhRvD.112l3054S}.

\begin{figure*}[ht]
\centering
\includegraphics[width=0.95\linewidth]{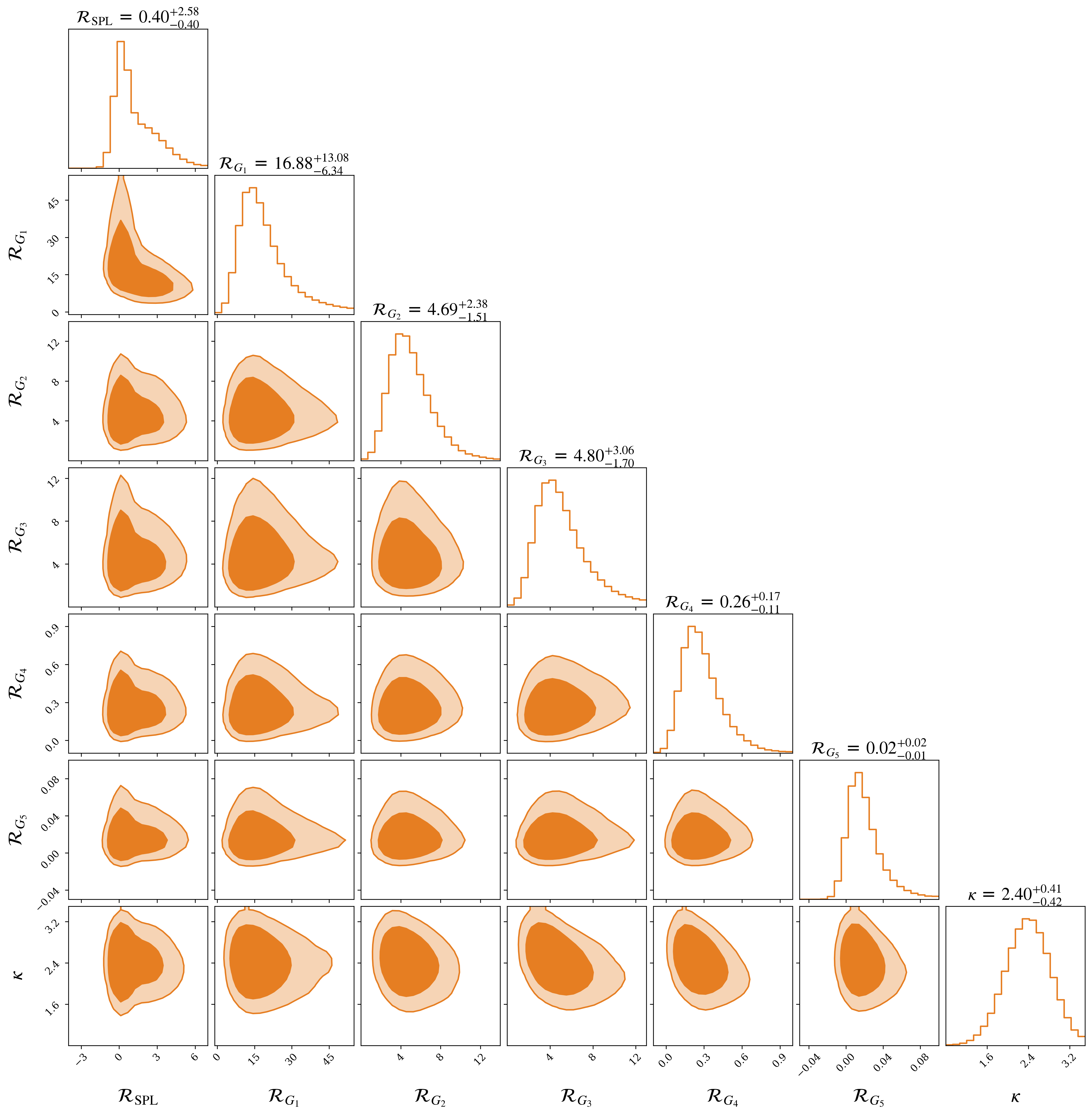}
\caption{This plot shows the inferred merger rate of each component, in units of ${\rm Gpc}^{-3}{\rm yr}^{-1}$. Rates are displayed on a linear scale. The contours enclose 68\% and 95\% of the posterior probability.}
\label{fig:rates_results}
\end{figure*}

\subsection{Redshift distribution}
\begin{figure}[ht]
\centering
\includegraphics[width=0.98\linewidth]{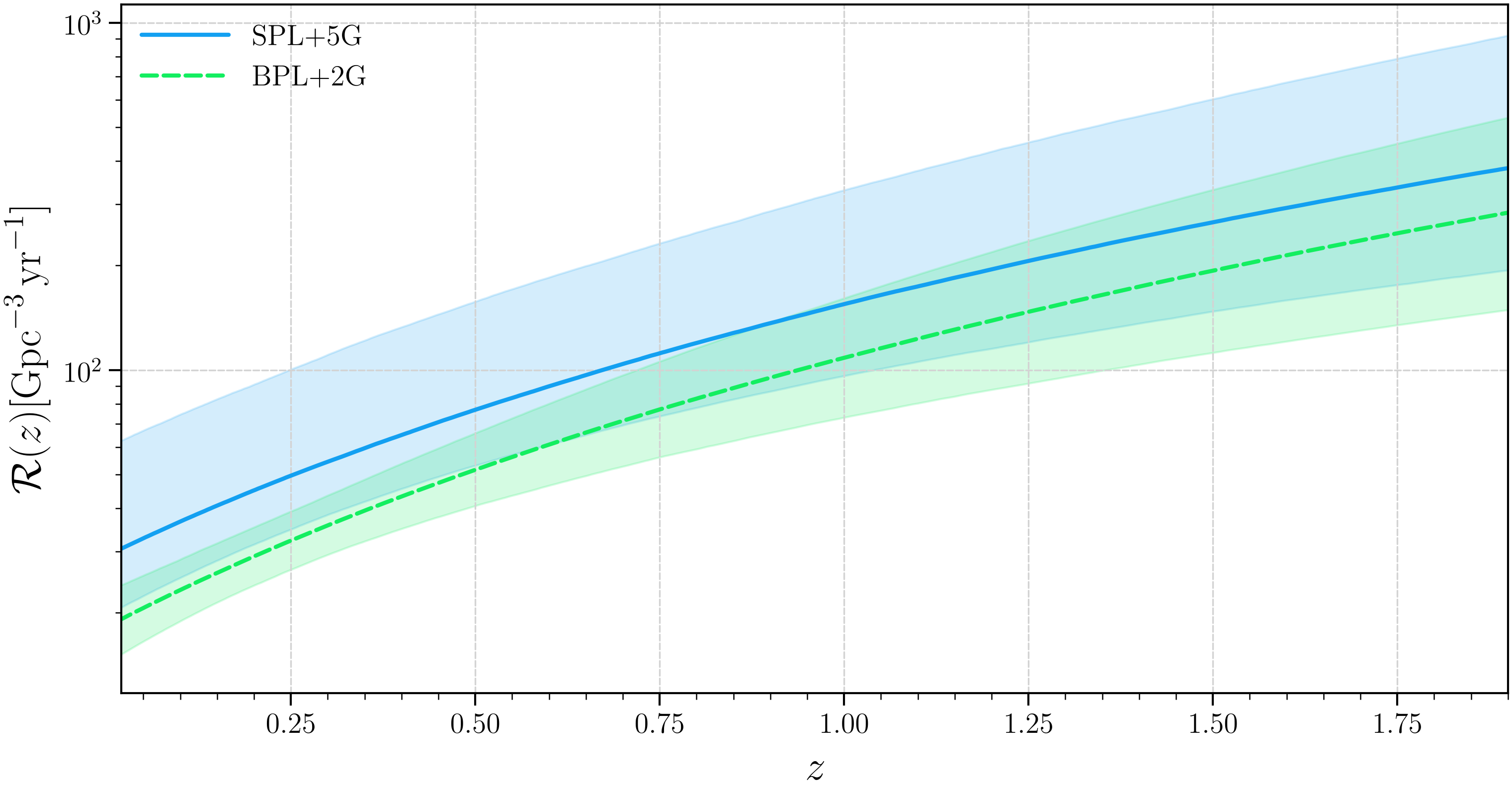}
\caption{Comparison of the inferred redshift distributions from the default O4b broken-power-law plus two-Gaussian (BPL+2G) model and our smoothed-power-law plus five-Gaussian (SPL+5G) model.}
\label{fig:redshift_dist}
\end{figure}

Figure \ref{fig:redshift_dist} shows the recovered redshift distributions obtained with our model and the default O4b
model. The two results are consistent.

\subsection{Single-component merger rates and properties}

Figure \ref{fig:rates_results} and Table \ref{tab:Rates} report the net volumetric  merger rates associated with the six
components of our mixture model.  Table  \ref{tab:Rates} also summarizes the characteristic features of each component,
as described above.
The high merger rate of $G_1$ could provide enough remnants to seed subsequent generations, provided they are retained.
The similarity between $G_2$ and $G_3$, as well as the mass cascade, suggests a common origin and near-equilibrium
merger rate.    Conversely, the cascade from $G_3\rightarrow G_5$ more closely resembles the ``starved'' hierarchical
cascades that occur in very inefficient hierarchical environments  \cite{gwastro-PopulationReconstruct-Hierarchical-WysockiDoctor2019}.
The characteristic primary masses increase along the sequence: $G_2$ has masses compatible with remnants of $G_1$
mergers, and $G_3$ with remnants of $G_2$ mergers. However, the broad peaks overlap, so multiple
channels could contribute to each component.

\begin{table*}
\caption{\label{tab:Rates}
Summary of the inferred local merger rates and characteristic properties of the components in the multi-component population model.  The quoted values are approximate posterior-median scales inferred from the component distributions.  Rates are reported in units of ${\rm Gpc}^{-3}{\rm yr}^{-1}$.  The final component is less well constrained because its support extends into the high-mass tail of the model.
}
\begin{ruledtabular}
\begin{tabular}{lcccc}
Component
& $R_k$
& $m_1$ scale
& $m_2$ scale
& Spin trend \\
\hline
SPL
& $\sim 1.56^{+2.03}_{-1.51}$
& $\sim 6\,M_\odot$
& $\sim 5\,M_\odot$
& very low spin \\
$G_1$
& $\sim 15.6^{+13.25}_{-5.48}$
& $\sim 8.5$--$9\,M_\odot$
& $\sim 7$--$8\,M_\odot$
& low spin, aligned \\
$G_2$
& $\sim 4.89^{+2.39}_{-1.62}$
& $\sim 16$--$17\,M_\odot$
& $\sim 13\,M_\odot$
& moderate spin, isotropic tilt \\
$G_3$
& $\sim 4.91^{+2.89}_{-1.68}$
& $\sim 27$--$32\,M_\odot$
& $\sim 31$--$32\,M_\odot$
& low/moderate spin, isotropic tilt \\
$G_4$
& $\sim 0.27^{+0.19}_{-0.11}$
& $\sim 45$--$55\,M_\odot$
& $\sim 45\,M_\odot$
& broad, higher spin \\
$G_5$
& $\sim 0.02^{+0.02}_{-0.01}$
& $\sim 80$--$85\,M_\odot$
& $\sim 80$--$85\,M_\odot$
& high spin, uncertain tail \\
\end{tabular}
\end{ruledtabular}
\end{table*}


\section{Discussion}
\label{sec:discuss}

The inferred mass structure admits a hierarchical interpretation, but the spin trends are in tension with this
interpretation.  In this section, we speculate about the self-consistent properties required for a putative phenomenological
multistage hierarchical formation
scenario that could explain the full population.
If some (or all) of the subpopulations apparent in the GW census are interrelated through a common hierarchical origin,
then their merger rates and properties should be related.
A unified hierarchical formation model must explain (a) the masses, (b) the surprising structure in the merger rates
versus mass, and (c) the absence of distinctively hierarchical spins.

\subsection{Brief review of hierarchical formation}
In this section, we briefly review a conventional phenomenological coagulation model for hierarchical formation, following Doctor et al. 
\cite{gwastro-PopulationReconstruct-Hierarchical-WysockiDoctor2019}.
For any given source environment, we characterize the merger, loss, and seeding of black holes by a continuous-time
coagulation equation of the form~\cite{Smoluchowski1916}
\begin{align}
\partial_t f(x;t) ={}&
\frac{1}{2}
\int dx'\,dx''\,
f(x';t) f(x'';t)
\Gamma(x',x'';t) \nonumber\\
&\times
\delta\!\left[
  x_{\rm rem}(x',x'') - x
\right] \nonumber\\
&-
\int dx'\,
f(x;t) f(x';t)
\Gamma(x,x';t) \nonumber\\
&+ r(x;t) - d(x;t) .
\label{eq:coag}
\end{align}
Here $x = (m, \chi)$ denotes single-black-hole parameters, $f(x;t)$ is the
black-hole distribution function, $\Gamma(x,x';t)$ is a volume-averaged
interaction rate, $x_{\rm rem}(x,x')$ maps merging components to remnant
parameters, and $r$ and $d$ are augmentation and depletion rates. The first
integral accumulates remnants; the second removes merged progenitors; the
integrand $f f' \Gamma$ of the second term \emph{is} the merger rate density
over parameters — the observable this model predicts.
The work by Doctor et al. assumed a simple universal pairing model, appropriate to a common formation environment, with
parametric form
\begin{equation}
  K_{\rm pair}(m_1, m_2) \propto (m_1 + m_2)^{a}\, q^{b},
  \qquad q \le 1,
  \label{eq:kernel}
\end{equation}
where the total-mass term represents interaction cross sections and dynamical
friction (gravitationally focused spheres would give $a = 2$), and the
mass-ratio term represents a preference for equal- or unequal-mass pairings (e.g.\ mass
segregation in clusters favors comparable masses).
Physically, we would anticipate the interaction rate in any one environment to increase with mass.

\subsection{Seeded, starved hierarchical formation}
In our case, the coagulation equation can be approximately reformulated as a discrete sum over distinct subpopulations with
number count $N_x$.
The behavior of the solutions depends on the seeding rate and the duration of the interaction phase. With steady seeding from an external
source, lower-generation mergers may reach rate equilibrium 
(i.e., $\Gamma_{a} N_a^2 = \Gamma_b N_b^2$ to stabilize the abundance $N_b$), such that each generation has the same
merger rate.  Qualitatively,  this rate-equilibrium is consistent with $R_{G_2}=R_{G_3}$.    For short interaction
times, the inherent rarity of sufficiently high generation mergers means that very high-generation mergers are starved of
progenitors, strongly suppressing high-generation mergers.  The rapid decrease in merger rate at high mass is
qualitatively consistent with such a constrained cascade.

The single-component merger rates provide a natural mechanism to assess hierarchical formation scenarios directly.  As
noted above, a model in which low-mass seeds are supplied at a steady rate and form binaries naturally explains the
first peak ($G_1$).  If the remnants are retained with high efficiency and propagated to another environment with a
sufficiently high interaction rate, they will naturally produce binaries with the masses and event rates associated with
$G_2$ and $G_3$. However, in order to prevent overproducing binaries with even higher masses, the interaction rate and
duration must suppress higher generation mergers.  

To assess the self-consistency of high-mass mergers as a ``starved'' hierarchical cascade,  we assume  $R_{c} = K R_{a}R_b$ where $K $ has
units of volume multiplied by time.
Table \ref{tab:EffectiveRates} reports this factor, for different combinations of transition paths.
For the higher-generation mergers, we find these factors are roughly consistent with one another, suggesting a similar
origin as starved hierarchical mergers.

\begin{table}[t]
\caption{\label{tab:mass_ladder}
Posterior-inferred characteristic primary-mass scales of the Gaussian
subpopulations. Peak masses are obtained from the maxima of the posterior
rate-density distributions. The final column reports the ratio of adjacent
peak masses, quantifying the spacing of the inferred mass ladder. Quoted
uncertainties correspond to central 90\% credible intervals.
}
\begin{ruledtabular}
\begin{tabular}{lcc}
Component &
$m_{\rm peak}\,[M_\odot]$ &
$m_{\rm peak}(G_k)/m_{\rm peak}(G_{k-1})$ \\ \hline
$G_1$ &
$9.56_{-2.78}^{+0.99}$ &
--- \\
$G_2$ &
$17.53_{-3.98}^{+3.18}$ &
$1.86_{-0.45}^{+0.76}$ \\
$G_3$ &
$27.69_{-2.39}^{+4.18}$ &
$1.59_{-0.29}^{+0.51}$ \\
$G_4$ &
$52.79_{-6.77}^{+7.76}$ &
$1.89_{-0.34}^{+0.37}$ \\
$G_5$ &
$84.46_{-13.34}^{+5.17}$ &
$1.57_{-0.28}^{+0.29}$ \\
\end{tabular}
\end{ruledtabular}
\end{table}

\begin{table*}
\caption{\label{tab:EffectiveRates}
Phenomenological effective interaction four-volumes inferred from the hyperposterior samples of component rates, reported as medians and 90\% credible intervals.  These quantities are bookkeeping factors used to compare possible hierarchical transition paths, not literal microscopic cross sections.  Values of $K_{\rm eff}$ are reported in units of ${\rm Gpc}^{3}{\rm yr}$.
}
\begin{ruledtabular}
\begin{tabular}{lcc}
Channel
& Definition
& $K_{\rm eff}$ \\
\hline
$G_1 + G_1 \rightarrow G_2$
& $R_{G_2}/R_{G_1}^2$
& $\sim 0.81^{+0.30}_{-0.32}$ \\
$G_1 + G_2 \rightarrow G_3$
& $R_{G_3}/(R_{G_1}R_{G_2})$
& $\sim 1.24^{+0.48}_{-0.75}$ \\
$G_2 + G_2 \rightarrow G_3$
& $R_{G_3}/R_{G_2}^2$
& $\sim 1.33^{+0.79}_{-0.52}$ \\
$G_2 + G_2 \rightarrow G_4$
& $R_{G_4}/R_{G_2}^2$
& $\sim 1.82^{+0.62}_{-0.79}$ \\
$G_2 + G_3 \rightarrow G_4$
& $R_{G_4}/(R_{G_2}R_{G_3})$
& $\sim 1.19^{+0.79}_{-0.54}$ \\
$G_3 + G_3 \rightarrow G_5$
& $R_{G_5}/R_{G_3}^2$
& $\sim 0.38^{+0.53}_{-0.21}$ \\
$G_3 + G_4 \rightarrow G_5$
& $R_{G_5}/(R_{G_3}R_{G_4})$
& $\sim 0.28^{+0.24}_{-0.11}$ \\
\end{tabular}
\end{ruledtabular}
\end{table*}


\subsection{An illustrative, finely tuned spin-down model}
In this section, we sketch a straw-man scenario to produce the desired spin structure. Roughly speaking, our model is
only consistent with hierarchical formation  if every binary black hole with mass
$M\le 50 M_\odot$ is either born with low natal spin (in $G_1$) or spins down rapidly.  Axion clouds provide a natural mechanism
to spin down black holes \cite{2015PhRvD..91h4011A,2011PhRvD..83d4026A,2019PhRvD..99j3015T,2026PhRvD.113h3031C}.
Axion spin-down acts extremely rapidly, with spin-down rates  $M^{-1}\chi  (M/\lambda_a)^9/24 $ for an axion with  Compton
wavelength $\lambda_a$, causing the black  hole spin to approach an attractor.  However, the superradiant axion
instability window reduces the BH spin only until the superradiance condition is no longer satisfied.
Expressed in terms of $\alpha(M,m_a) = GMm_a/\hbar c$, spin-down terminates when
\begin{equation}
  \omega_{n\ell m} \simeq m\Omega_H .
\end{equation}
As a result, axion spin-down converges to a mass-dependent attractor; roughly speaking, using a hydrogen-atom
approximation for the natural axion eigenfrequencies
\begin{equation}
  \omega_{n\ell m}
  \simeq
  \mu_a\left(1-\frac{\alpha^2}{2n^2}\right),
\end{equation}
the Regge curve associated with a specific level is
\begin{equation}
  \chi_{n\ell m}
  \simeq
  \frac{4k_{nm}}{1+4k_{nm}^2},
  \qquad
  k_{nm}
  \equiv
  \frac{\alpha}{m}
  \left(1-\frac{\alpha^2}{2n^2}\right).
  \label{eq:multi-regge-spin}
\end{equation}
Because multiple transitions can contribute, the limiting spin can be quite small across the whole mass spectrum.
However, superradiant instability only progresses if the axion cloud persists. 
Axion clouds can be disrupted by dynamical interactions; in an AGN disk, massive black holes will frequently encounter
neighboring  black holes or nearby stars
\cite{gwastro-agndisk-McFacts1-Core-2024,gwastro-agndisk-McFacts3-Populations-2024,gwastro-agndisk-VeraGW231123McFacts,gwastro-agndisk-McFacts-NathanielThesisPhD}.

\begin{figure}
\includegraphics[width=\columnwidth]{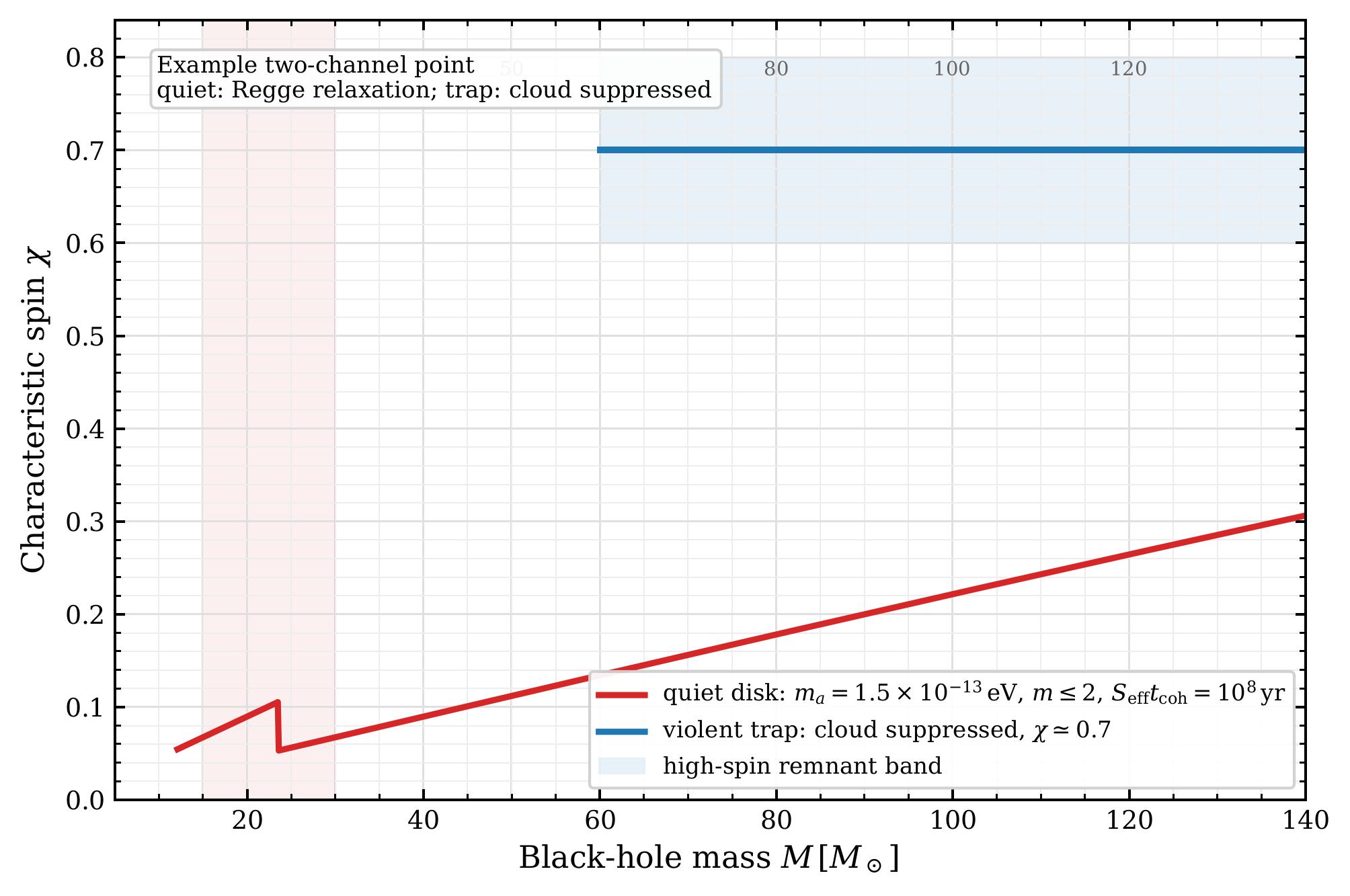}
\caption{\label{fig:spin_axion_cartoon}Illustrative model of black-hole spin-down and high-mass spin retention}
\end{figure}

Motivated by these two processes, we propose an extreme straw-man model: axionic spin-down, inhibited by occasional rapid encounters at high mass.  This model would permit high-spin BH merger
remnants to retain their spin temporarily. Figure \ref{fig:spin_axion_cartoon} qualitatively illustrates the two-component
model's expectations for spin versus mass.  As outlined in Appendix \ref{ap:toy_model}, this model predicts a
potentially large but
not yet fully excluded stochastic GW background from these boson clouds.

\begin{figure*}
\includegraphics[width=\textwidth]{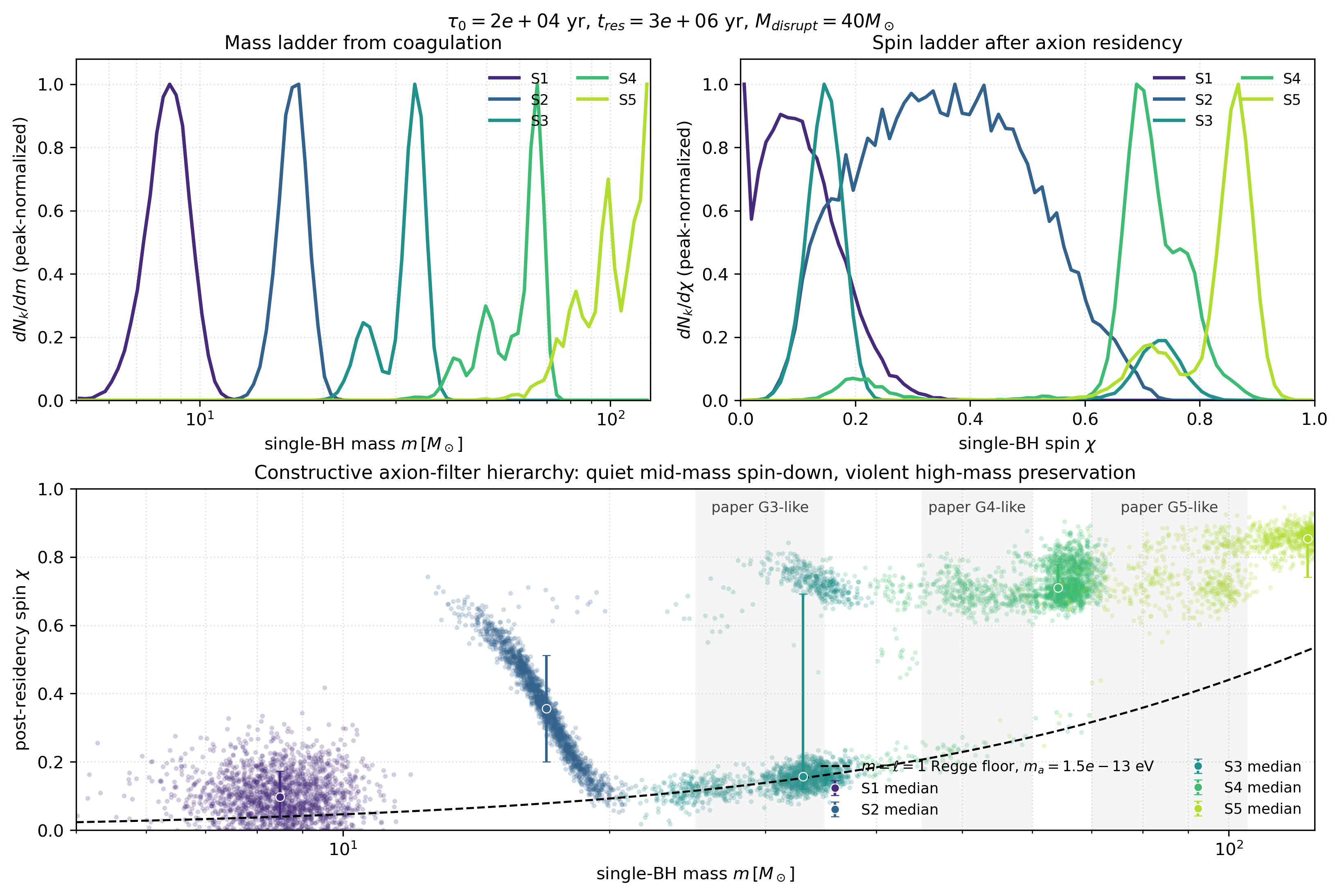}
\caption{\label{fig:full_hierarchy_consistency} Monte Carlo simulation of hierarchical BH formation  with axionic spin-down.
\emph{Top left panel}: Mass distribution of individual BHs.  In this figure, S1$\ldots$S5 refer to the specific
generations of BHs simulated in this model, not necessarily the \emph{joint} populations G1$\ldots$G5 recovered in our hierarchical
fit to GWTC-5.
\emph{Top right panel}: Spin distribution of individual BHs, after axion spin-down.
\emph{Bottom panel}: Joint mass and spin distribution of isolated BHs.  
}
\end{figure*}

\begin{figure*}
\includegraphics[width=\textwidth]{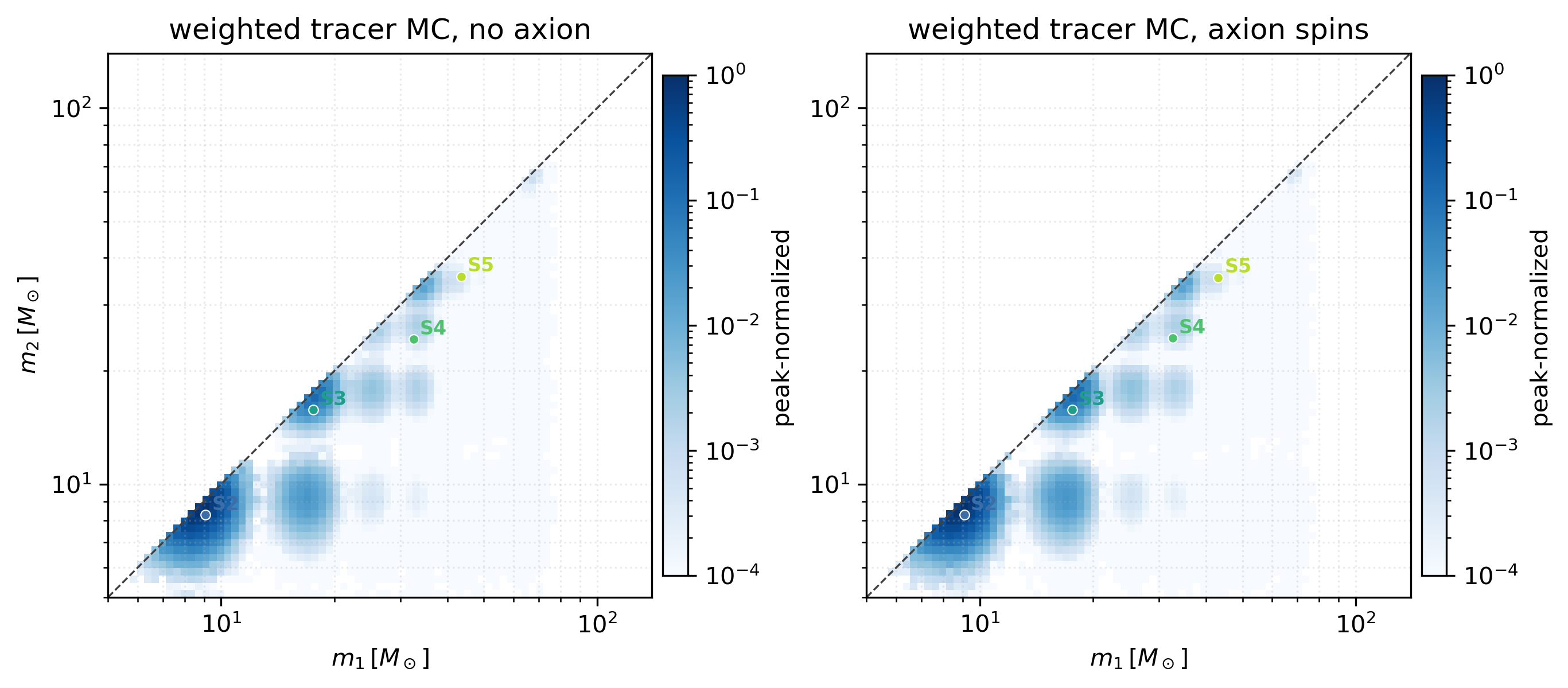}
\caption{\label{fig:full_hierarchy_2d} Binary merger rate versus component masses expected from a Monte Carlo simulation
of hierarchical binary formation, including axionic spin-down. 
}
\end{figure*}

Figure \ref{fig:full_hierarchy_consistency} shows the predictions of a single Monte Carlo simulation using our coagulation
model, modified to account for black hole spin-down due to axions.   The parameters of this simulation have been
hand-tuned to qualitatively reproduce the features seen in our population model.  Figure \ref{fig:full_hierarchy_2d}
shows the associated two-dimensional merger rate versus component masses.   The simulation framework is described in companion work and outlined in Appendix~\ref{ap:constructive_model}.

\section{Conclusion}
\label{sec:conclude}

In this paper, we fit a multi-component mixture  model to the binary black hole census through  GWTC-5 to test the
spin consistency of a hierarchical interpretation of previously identified mass structure \cite{2026arXiv260414290G}.
The adopted mass hierarchy is an ansatz, motivated by and building upon previous analyses which provide robust evidence for structure
compatible with a hierarchical interpretation.
Conditional on this ansatz, our recovered spin distribution is in tension with a naive hierarchical interpretation: black hole spins
from ``merger remnant'' populations are not equal to the expected value predicted by vacuum general relativity.  Our
population also differs from previous phenomenological analyses that incorporated this post-merger spin
assumption: unlike previous work, only the lowest-mass Gaussian component favors aligned spins, while the higher-mass components are consistent with isotropic orientations.
Our spin trends in orientation versus mass seem qualitatively consistent with other recent
work \cite{2026arXiv260614472F} which favors a symmetric $\chi_{\rm eff}$ distribution for many mass bins, with
extensions toward positive $\chi_{\rm eff}$. 

The puzzling spins remain the single most difficult feature to explain within a conventional hierarchical formation
scenario.  We propose a fine-tuned straw-man model to explain the observed trends of spin versus mass: axionic
spin-down globally suppressing spins for almost all black holes, limited by rare rapid interactions at high mass. This finely tuned model makes extremely strong predictions, since most merging BHs would be forming
hierarchically and thus inherit large spins, which are promptly lost to axions.  In particular, it would predict a
stochastic background closely tied to the axion mass scale
\begin{equation}
  f_{\rm GW}
  \simeq \frac{2m_a c^2}{h}
  \simeq 48.4\,{\rm Hz}
  \left(\frac{m_a}{10^{-13}\,{\rm eV}}\right).
\end{equation}
The large predicted stochastic background is in tension with current observational limits, suggesting the most naive
version of this model could be ruled out completely in the next few years.




\section*{Acknowledgements}
This material is based upon work supported by the NSF's LIGO Laboratory, a major facility fully funded by the National Science Foundation. The authors acknowledge the computational resources provided by the LIGO Laboratory's CIT cluster, which is supported by National Science Foundation Grants PHY-0757058 and PHY0823459. ROS acknowledges support from NSF Grant No. AST-1909534, NSF Grant No. PHY-2012057, and the Simons Foundation.


\appendix

\section{Constructive model for hierarchical formation}
\label{ap:constructive_model}

This appendix describes the forward model used to generate the constructive
hierarchical examples.  The goal is not to define a final population model,
but to give a self-consistent toy interaction prescription which produces a
mass hierarchy while allowing axion spin filtering to act on some remnants.
Throughout this appendix we use $S_1,S_2,\ldots$ to denote
single-black-hole/remnant generations.  This notation is deliberately distinct
from the $G_1,\ldots,G_5$ binary mixture components used in the main
population fit.
The phenomenological hierarchical interaction model is generic; however, the many finely tuned choices invoked in the
interaction model and axion model were developed, documented, and simulated with an AI system (Codex).

\subsection{Tracer populations and abundance weights}

The Monte Carlo implementation uses tracer particles for the shape of each
single-black-hole generation, but it carries the scalar abundance of each
generation separately.  We denote the normalized tracer distribution by $p_k(m,\chi)$ 
for single black holes in generation
$S_k$, and let $N_k$ denote its scalar abundance.
The initial population is a
truncated Gaussian in mass and spin,
\begin{align}
  m &\sim {\cal N}_{[m_{\min},m_{\max}]}(\mu_m,\sigma_m),\\
  \chi &\sim {\cal N}_{[0,0.98]}(\mu_\chi,\sigma_\chi),
\end{align}
with the default values
$  \mu_m=8.5M_\odot$,
$ \sigma_m=1.0M_\odot$, 
$m_{\min}=4M_\odot$,
$ m_{\max}=140M_\odot$, with spin parameters
$\mu_\chi=0.08$ and $\sigma_\chi=0.08$.
We set $N_1=1$; all later $N_k$ are generated by the interaction model
below.  In the code, each generation is represented by a fixed number of
tracer particles, but this fixed tracer count is only a shape estimator.  It
is not used as the physical abundance.

\subsection{Generation recurrence}

The finite-generation recurrence follows the Doctor-style coagulation
construction used in the demonstration code: to build generation $S_k$, the
newest available generation $S_{k-1}$ is paired against all earlier resident
generations $S_j$ with $1\leq j\leq k-1$.  Thus
\begin{equation}
  S_k \leftarrow S_j + S_{k-1},
  \qquad j=1,\ldots,k-1 .
\end{equation}
For each channel $(j,\ell)$, with $\ell=k-1$, the differential interaction
weight is
\begin{align}
dR_{j\ell}
={}&
\epsilon_{\rm int}\,
c_{j\ell}\,
N_j N_\ell\,
K(m_1,m_2)
\nonumber\\
&\times
p_j(m_1,\chi_1)\,
p_\ell(m_2,\chi_2)\,
dm_1\,d\chi_1\,dm_2\,d\chi_2 ,
\label{eq:app-channel-rate}
\end{align}

where

\begin{equation}
  c_{j\ell} =
  \begin{cases}
  1/2, & j=\ell,\\
  1, & j\neq \ell ,
  \end{cases}
\end{equation}
and $\epsilon_{\rm int}$ is an effective interaction four-volume.  The
default used in the demonstration is
\begin{equation}
  \epsilon_{\rm int}=3.0\times10^{-4}.
\end{equation}
This factor is the explicit prefactor that prevents the high-generation
cascade from being artificially amplified by the fixed tracer count.  Without
it, a fixed-size tracer cloud can make high-generation mergers look far more
common than they are.

The remnant of each retained merger is mapped to a new mass and birth spin
using the remnant fits already implemented in the population engine.  The
retained contribution to the next generation is then
\begin{equation}
  N_k =
  \sum_{j=1}^{k-1}
  \int dR_{j,k-1}\,
  P_{\rm ret}(m_1,m_2,\chi_1,\chi_2),
  \label{eq:app-generation-abundance}
\end{equation}
where the demonstration used a high-retention environment,
 $v_{\rm esc}=1000~{\rm km\,s^{-1}}$.
The normalized shape $p_k(m,\chi)$ is obtained by resampling the retained
remnant pool with weights proportional to the integrand in
Eq.~(\ref{eq:app-generation-abundance}).  This separates the tracer shape
from the physical channel abundance.

\subsection{High-mass locality kernel}

A separable power-law pairing kernel,
\begin{equation}
  K_{\rm sep}(m_1,m_2)\propto (m_1+m_2)^a q^b,
  \qquad
  q=\frac{\min(m_1,m_2)}{\max(m_1,m_2)},
\end{equation}
does not by itself suppress the abundant mixed channels involving $S_1$.
Even with a large global $q$ exponent, the reservoir abundance $N_1$ can make
$S_1+S_3$ or $S_1+S_4$ dominate the high-mass event budget.  The constructive
model therefore uses a mass-dependent locality factor,
\begin{equation}
  K(m_1,m_2)
  =
  (m_1+m_2)^a q^{b_{\rm low}}\,
  {\cal L}(m_1,m_2),
  \label{eq:app-locality-kernel}
\end{equation}
where
\begin{equation}
  {\cal L}
  =
  (1-H)
  +
  H\,A_{\rm loc}
  \exp\left[
    -\frac{1}{2}
    \left(
      \frac{\ln(m_{\rm big}/m_{\rm small})}{\sigma_{\rm eff}}
    \right)^2
  \right].
  \label{eq:app-locality-factor}
\end{equation}
Here 
  $m_{\rm big}=\max(m_1,m_2),
  m_{\rm small}=\min(m_1,m_2)$
and the high-mass switch is
\begin{equation}
  H(M_{\rm tot})
  =
  \left[
  1+
  \exp\left(
    -\frac{M_{\rm tot}-M_{\rm turn}}{\Delta M_{\rm turn}}
  \right)
  \right]^{-1},
\end{equation}
with $M_{\rm tot}=m_1+m_2$. The width of the log-mass locality window narrows with total mass,
\begin{equation}
  \sigma_{\rm eff}
  =
  \sigma_{\rm low}
  +
  (\sigma_{\rm high}-\sigma_{\rm low})H .
\end{equation}

Table \ref{tab:locality_kernel_params} shows the parameter values used for the default comparison.
The large value of $b_{\rm low}$ reduces low-mass unequal-mass
encounters without eliminating $S_1+S_2$ growth.  The high-mass locality
factor then suppresses mixed channels such as $S_1+S_4$ and $S_2+S_4$ unless
the masses are comparable.  Physically, this is meant to represent a dense
AGN-disk or migration-trap environment in which the most massive objects
interact preferentially with comparable-mass objects in the same local
dynamical substructure, rather than with the entire low-mass reservoir.

\begin{table*}
\caption{\label{tab:locality_kernel_params}
Parameters adopted for the locality-enhanced pairing kernel and interaction model used in the illustrative hierarchical-formation simulation.
}
\begin{ruledtabular}
\begin{tabular}{lcl}
Parameter & Value & Description \\
\hline
$a$ & $2.2$ & Total-mass enhancement exponent \\
$b_{\rm low}$ & $4.2$ & Low-mass equal-mass preference exponent \\
$M_{\rm turn}$ & $42\,M_\odot$ & Onset of high-mass locality \\
$\Delta M_{\rm turn}$ & $7\,M_\odot$ & Width of locality turn-on \\
$\sigma_{\rm low}$ & $0.9$ & Low-mass log-ratio width \\
$\sigma_{\rm high}$ & $0.28$ & High-mass log-ratio width \\
$A_{\rm loc}$ & $14$ & High-mass locality enhancement \\
$\epsilon_{\rm int}$ & $3.0\times10^{-4}$ & Effective interaction four-volume \\
\end{tabular}
\end{ruledtabular}
\end{table*}

\subsection{Axion spin map}

The interaction model above determines the mass-pairing and channel weights.
The axion model is applied as a post-merger spin-residency map for retained
single black holes.  The Regge floor is approximated by the leading
$m=\ell=1$ scalar trajectory,
\begin{equation}
  \alpha(M;m_a)
  =
  0.0748
  \left(\frac{M}{10M_\odot}\right)
  \left(\frac{m_a}{10^{-12}\,{\rm eV}}\right),
\end{equation}
\begin{equation}
  \chi_{\rm R}(M;m_a)
  =
  \frac{4\alpha}{1+4\alpha^2}.
\end{equation}
The post-residency spin is
\begin{equation}
  \chi_{\rm out}
  =
  \chi_{\rm R}
  +
  (\chi_{\rm birth}-\chi_{\rm R})\exp(-\Lambda_{\rm eff}),
  \label{eq:app-axion-map}
\end{equation}
with
\begin{equation}
  \Lambda_{\rm eff}=\frac{t_{\rm coh}}{\tau_{\rm ax}(M)} .
\end{equation}
The stimulated spin-down time used in the toy model is
\begin{equation}
  \tau_{\rm ax}(M)
  =
  \tau_0
  \left[
    \frac{\alpha(M;m_a)}{\alpha_{\rm ref}}
  \right]^{-p_\tau}.
\end{equation}
\begin{table*}
\caption{\label{tab:axion_residency_params}
Parameters adopted for the illustrative axion spin-filtering and residency model used in the hierarchical-formation simulation.
}
\begin{ruledtabular}
\begin{tabular}{lcl}
Parameter & Value & Description \\
\hline
$m_a$ & $1.55\times10^{-13}\,{\rm eV}$ & Scalar mass \\
$\tau_0$ & $2.0\times10^4\,{\rm yr}$ & Reference spin-down time \\
$\alpha_{\rm ref}$ & $0.035$ & Reference coupling \\
$p_{\tau}$ & $9$ & Growth-time scaling exponent \\
$t_{\rm res}$ & $3.0\times10^6\,{\rm yr}$ & Quiet residence time \\
$M_{\rm disrupt}$ & $40\,M_\odot$ & Onset of violent-channel cloud disruption \\
$\Delta M_{\rm disrupt}$ & $5\,M_\odot$ & Disruption transition width \\
$\Lambda_{\rm cap}$ & $0.08$ & Maximum relaxation depth in disrupted channel \\
$t_{\rm enc,ref}$ & $2.0\times10^5\,{\rm yr}$ & Encounter time at $30\,M_\odot$ \\
$p_{\rm enc}$ & $3$ & Encounter-time mass exponent \\
$N_{\rm orb,min}$ & $4$ & Minimum coherent-orbit floor \\
$P_{\rm orb}$ & $10^3\,{\rm yr}$ & Fiducial SMBH orbital period \\
$\sigma_\chi$ & $0.025$ & Residual spin scatter \\
\end{tabular}
\end{ruledtabular}
\end{table*}
The probability of being in the quiet channel is
\begin{equation}
  p_{\rm quiet}(M)
  =
  \left[
  1+
  \exp\left(
    \frac{M-M_{\rm disrupt}}{\Delta M_{\rm disrupt}}
  \right)
  \right]^{-1}.
\end{equation}
Quiet objects use $t_{\rm coh}=t_{\rm res}$.  Disrupted objects instead use
\begin{equation}
  t_{\rm coh}
  =
  \max\left[
    t_{\rm enc,ref}\left(\frac{M}{30M_\odot}\right)^{-p_{\rm enc}},
    N_{\rm orb,min}P_{\rm orb}
  \right],
\end{equation}
and their relaxation depth is capped,
\begin{equation}
  \Lambda_{\rm eff}
  =
  \min\left(
    \frac{t_{\rm coh}}{\tau_{\rm ax}},
    \Lambda_{\rm cap}
  \right).
\end{equation}
This implements the intended phenomenology: quiet low/intermediate-mass
residents can spin down efficiently, while high-mass objects in a violent
trap-like environment retain much of their merger birth spin.

\section{Reconciling axion spin-down and LVK observations}
\label{ap:toy_model}
The axion spin-filtering scenario also predicts gravitational radiation from
the boson clouds produced during spin-down.  This radiation can be substantial in the proposed scenario, which assumes that most observed BH mergers
form hierarchically. Their high-spin remnants can transfer a substantial fraction of their mass-energy to boson clouds during rapid
spin-down. For a sense of scale, if a
black hole with initial spin $\chi_i$ relaxes to $\chi_f$, the available rotational energy is

\begin{align}
\frac{E_{\rm ext}}{c^2}
&=
M(\chi_i)-M(\chi_f),
\nonumber\\
M(\chi)
&=
\frac{\sqrt{2}\,M_{\rm irr}}
{\sqrt{1+\sqrt{1-\chi^2}}} .
\label{eq:extractable-energy}
\end{align}


For a typical hierarchical merger remnant with $\chi_i\simeq0.7$ relaxing
to a low Regge spin $\chi_f\simeq0.1$, the extracted energy is of order
$  E_{\rm ext}
  \sim
  0.07\,M c^2
  \sim
  (1\mbox{--}2)M_\odot c^2
$%
 for $M\simeq15$--$30M_\odot$.
If emitted as gravitational waves, this energy contributes to a stochastic background concentrated in frequency, with an approximate amplitude 
\begin{align}
\Omega_{\rm GW}
&\sim
\frac{R\,E_{\rm GW}}
     {\rho_c c^2 H_0}
\nonumber\\
&\simeq
1.1\times10^{-10}
\left(
\frac{R}
     {1\,{\rm Gpc}^{-3}{\rm yr}^{-1}}
\right)
\left(
\frac{E_{\rm GW}}
     {M_\odot c^2}
\right)
\xi_z ,
\label{eq:omega-budget}
\end{align}
where $\xi_z$ accounts for redshift evolution and for the difference between
the local and cosmological formation rate.  Taking the rough rate of the
lowest Gaussian component, $R_{\rm G1}\sim 11\,{\rm Gpc}^{-3}{\rm yr}^{-1}$,
and $E_{\rm GW}\sim1$--$1.5M_\odot c^2$ gives
\begin{equation}
  \Omega_{\rm GW}
  \sim
  (1.3\mbox{--}1.8)\times10^{-9}\,\xi_z .
\end{equation}
Thus the most naive implementation, in which every G1-rate hierarchical
remnant forms a long-lived cloud and radiates most of the extracted spin
energy into gravitational waves, can easily approach or exceed present LVK
stochastic limits.

This conclusion should be compared with the generic LVK limits on an
isotropic gravitational-wave background.  The O3 analysis gives limits at
the few $\times10^{-9}$ level \cite{LIGO-O3-stochastic-isotropic}, and the O4a
analysis gives
\begin{equation}
  \Omega_{\rm GW}(25\,{\rm Hz}) \leq 2.0\times10^{-9}
\end{equation}
for a compact-binary-like $f^{2/3}$ spectrum and
$\Omega_{\rm GW}\leq2.8\times10^{-9}$ for a flat spectrum
\cite{LIGO-O4a-stochastic-isotropic}.  Dedicated scalar-cloud reinterpretations of
O1--O3 data are even more directly relevant: Yuan, Jiang, and Huang find no
evidence for a scalar-cloud background and, depending on the assumed
stellar-origin black-hole spin distribution, exclude intervals such as
$[1.5,16]\times10^{-13}\,{\rm eV}$, $[1.9,8.3]\times10^{-13}\,{\rm eV}$,
and $[1.3,17]\times10^{-13}\,{\rm eV}$ at 95 percent credibility
\cite{2022PhRvD.106b3020Y}.  Spin-measurement constraints from GWTC-2 also
disfavor scalar masses in the same neighborhood under long-isolation
assumptions \cite{2021PhRvL.126o1102N}.
Therefore, the most naive version of this model is already at the limits of current observations. 

A dense AGN-disk environment provides many opportunities to disrupt the boson cloud and mitigate these extreme limits,
relaxing the requirement that all axionically-extracted energy be emitted through GW.  
Some fraction
 of the extracted energy will be lost  to
black-hole reabsorption, bosenovae, scalar radiation, environmental
ionization, or tidal depletion.  Dynamically active AGN migration traps
can reduce the gravitational-wave emission efficiency: close companions, eccentric
passages, binary-single encounters, gas torques, and tidal level mixing can
interrupt the cloud before it radiates efficiently.

\bibstyle{unsrt}
\bibliography{combined}

@ARTICLE{2026arXiv260618081T,
       author = {{Tiwari}, Vaibhav},
        title = "{The Chirp-Mass Ladder: A New Rung Emerges}",
      journal = {arXiv e-prints},
         year = 2026,
        month = jun,
          eid = {arXiv:2606.18081},
        pages = {arXiv:2606.18081},
          doi = {10.48550/arXiv.2606.18081},
archivePrefix = {arXiv},
       eprint = {2606.18081},
 primaryClass = {astro-ph.HE},
       adsurl = {https://ui.adsabs.harvard.edu/abs/2026arXiv260618081T}
}

@ARTICLE{2026arXiv260602318P,
       author = {{Padhyegurjar}, Shaunak and {Mukherjee}, Suvodip},
        title = "{The First Detection of Sub-Populations in the Delay-Time Distribution of Binary Black Holes in GWTC-4 of LIGO-Virgo-KAGRA}",
      journal = {arXiv e-prints},
         year = 2026,
        month = jun,
          eid = {arXiv:2606.02318},
        pages = {arXiv:2606.02318},
          doi = {10.48550/arXiv.2606.02318},
archivePrefix = {arXiv},
       eprint = {2606.02318},
 primaryClass = {astro-ph.HE},
       adsurl = {https://ui.adsabs.harvard.edu/abs/2026arXiv260602318P}
}

@Article{	  1995apjs...99..609s,
  author	= {{Sigurdsson}, Steinn and {Phinney}, E.~S.},
  title		= "{Dynamics and Interactions of Binaries and Neutron Stars
		  in Globular Clusters}",
  journal	= {\apjs},
  year		= 1995,
  month		= aug,
  volume	= {99},
  pages		= {609},
  doi		= {10.1086/192199},
  archiveprefix	= {arXiv},
  eprint	= {astro-ph/9412078},
  primaryclass	= {astro-ph},
  adsurl	= {https://ui.adsabs.harvard.edu/abs/1995ApJS...99..609S}
}

@Article{	  1998mnras.300..857e,
  author	= {{Elson}, Rebecca A.~W. and {Sigurdsson}, Steinn and
		  {Davies}, Melvyn and {Hurley}, Jarrod and {Gilmore},
		  Gerard},
  title		= "{The binary star population of the young cluster NGC 1818
		  in the Large Magellanic Cloud}",
  journal	= {\mnras},
  year		= 1998,
  month		= nov,
  volume	= {300},
  number	= {3},
  pages		= {857-862},
  doi		= {10.1046/j.1365-8711.1998.01948.x},
  adsurl	= {https://ui.adsabs.harvard.edu/abs/1998MNRAS.300..857E}
}

@InProceedings{	  2004aipc..735..195l,
  author	= {{Loredo}, Thomas J.},
  title		= {{Accounting for Source Uncertainties in Analyses of
		  Astronomical Survey Data}},
  booktitle	= {Bayesian Inference and Maximum Entropy Methods in Science
		  and Engineering: 24th International Workshop on Bayesian
		  Inference and Maximum Entropy Methods in Science and
		  Engineering},
  year		= 2004,
  editor	= {{Fischer}, Rainer and {Preuss}, Roland and {Toussaint},
		  Udo Von},
  series	= {American Institute of Physics Conference Series},
  volume	= {735},
  month		= nov,
  pages		= {195-206},
  doi		= {10.1063/1.1835214},
  archiveprefix	= {arXiv},
  eprint	= {astro-ph/0409387},
  primaryclass	= {astro-ph},
  adsurl	= {https://ui.adsabs.harvard.edu/abs/2004AIPC..735..195L}
}

@Article{	  2010cqgra..27k4007m,
  author	= {{Mandel}, I. and {O'Shaughnessy}, R.},
  title		= "{Compact binary coalescences in the band of ground-based
		  gravitational-wave detectors}",
  journal	= {Classical and Quantum Gravity},
  archiveprefix	= "arXiv",
  eprint	= {0912.1074},
  primaryclass	= "astro-ph.HE",
  year		= 2010,
  month		= jun,
  volume	= 27,
  number	= 11,
  pages		= {114007-+},
  doi		= {10.1088/0264-9381/27/11/114007},
  adsurl	= {http://adsabs.harvard.edu/abs/2010CQGra..27k4007M}
}

@Article{	  2011phrvd..83d4026a,
  author	= {{Arvanitaki}, Asimina and {Dubovsky}, Sergei},
  title		= "{Exploring the string axiverse with precision black hole
		  physics}",
  journal	= {\prd},
  year		= 2011,
  month		= feb,
  volume	= {83},
  number	= {4},
  eid		= {044026},
  pages		= {044026},
  doi		= {10.1103/PhysRevD.83.044026},
  archiveprefix	= {arXiv},
  eprint	= {1004.3558},
  primaryclass	= {hep-th},
  adsurl	= {https://ui.adsabs.harvard.edu/abs/2011PhRvD..83d4026A}
}

@Article{	  2015cqgra..32b4001a,
  author	= {{Acernese}, F. and {Agathos}, M. and {Agatsuma}, K. and
		  {Aisa}, D. and {Allemandou}, N. and {Allocca}, A. and
		  {Amarni}, J. and {Astone}, P. and {Balestri}, G. and
		  {Ballardin}, G. and {Barone}, F. and {Baronick}, J.-P. and
		  {Barsuglia}, M. and {Basti}, A. and {Basti}, F. and
		  {Bauer}, Th S. and {Bavigadda}, V. and {Bejger}, M. and
		  {Beker}, M.~G. and {Belczynski}, C. and {Bersanetti}, D.
		  and {Bertolini}, A. and {Bitossi}, M. and {Bizouard}, M.~A.
		  and {Bloemen}, S. and {Blom}, M. and {Boer}, M. and
		  {Bogaert}, G. and {Bondi}, D. and {Bondu}, F. and
		  {Bonelli}, L. and {Bonnand}, R. and {Boschi}, V. and
		  {Bosi}, L. and {Bouedo}, T. and {Bradaschia}, C. and
		  {Branchesi}, M. and {Briant}, T. and {Brillet}, A. and
		  {Brisson}, V. and {Bulik}, T. and {Bulten}, H.~J. and
		  {Buskulic}, D. and {Buy}, C. and {Cagnoli}, G. and
		  {Calloni}, E. and {Campeggi}, C. and {Canuel}, B. and
		  {Carbognani}, F. and {Cavalier}, F. and {Cavalieri}, R. and
		  {Cella}, G. and {Cesarini}, E. and {Mottin}, E. Chassande-
		  and {Chincarini}, A. and {Chiummo}, A. and {Chua}, S. and
		  {Cleva}, F. and {Coccia}, E. and {Cohadon}, P.-F. and
		  {Colla}, A. and {Colombini}, M. and {Conte}, A. and
		  {Coulon}, J.-P. and {Cuoco}, E. and {Dalmaz}, A. and
		  {D'Antonio}, S. and {Dattilo}, V. and {Davier}, M. and
		  {Day}, R. and {Debreczeni}, G. and {Degallaix}, J. and
		  {Del{\'e}glise}, S. and {Pozzo}, W. Del and {Dereli}, H.
		  and {Rosa}, R. De and {Fiore}, L. Di and {Lieto}, A. Di and
		  {Virgilio}, A. Di and {Doets}, M. and {Dolique}, V. and
		  {Drago}, M. and {Ducrot}, M. and {Endr{\H{o}}czi}, G. and
		  {Fafone}, V. and {Farinon}, S. and {Ferrante}, I. and
		  {Ferrini}, F. and {Fidecaro}, F. and {Fiori}, I. and
		  {Flaminio}, R. and {Fournier}, J.-D. and {Franco}, S. and
		  {Frasca}, S. and {Frasconi}, F. and {Gammaitoni}, L. and
		  {Garufi}, F. and {Gaspard}, M. and {Gatto}, A. and {Gemme},
		  G. and {Gendre}, B. and {Genin}, E. and {Gennai}, A. and
		  {Ghosh}, S. and {Giacobone}, L. and {Giazotto}, A. and
		  {Gouaty}, R. and {Granata}, M. and {Greco}, G. and {Groot},
		  P. and {Guidi}, G.~M. and {Harms}, J. and {Heidmann}, A.
		  and {Heitmann}, H. and {Hello}, P. and {Hemming}, G. and
		  {Hennes}, E. and {Hofman}, D. and {Jaranowski}, P. and
		  {Jonker}, R.~J.~G. and {Kasprzack}, M. and
		  {K{\'e}f{\'e}lian}, F. and {Kowalska}, I. and {Kraan}, M.
		  and {Kr{\'o}lak}, A. and {Kutynia}, A. and {Lazzaro}, C.
		  and {Leonardi}, M. and {Leroy}, N. and {Letendre}, N. and
		  {Li}, T.~G.~F. and {Lieunard}, B. and {Lorenzini}, M. and
		  {Loriette}, V. and {Losurdo}, G. and {Magazz{\`u}}, C. and
		  {Majorana}, E. and {Maksimovic}, I. and {Malvezzi}, V. and
		  {Man}, N. and {Mangano}, V. and {Mantovani}, M. and
		  {Marchesoni}, F. and {Marion}, F. and {Marque}, J. and
		  {Martelli}, F. and {Martellini}, L. and {Masserot}, A. and
		  {Meacher}, D. and {Meidam}, J. and {Mezzani}, F. and
		  {Michel}, C. and {Milano}, L. and {Minenkov}, Y. and
		  {Moggi}, A. and {Mohan}, M. and {Montani}, M. and
		  {Morgado}, N. and {Mours}, B. and {Mul}, F. and {Nagy},
		  M.~F. and {Nardecchia}, I. and {Naticchioni}, L. and
		  {Nelemans}, G. and {Neri}, I. and {Neri}, M. and {Nocera},
		  F. and {Pacaud}, E. and {Palomba}, C. and {Paoletti}, F.
		  and {Paoli}, A. and {Pasqualetti}, A. and {Passaquieti}, R.
		  and {Passuello}, D. and {Perciballi}, M. and {Petit}, S.
		  and {Pichot}, M. and {Piergiovanni}, F. and {Pillant}, G.
		  and {Piluso}, A. and {Pinard}, L. and {Poggiani}, R. and
		  {Prijatelj}, M. and {Prodi}, G.~A. and {Punturo}, M. and
		  {Puppo}, P. and {Rabeling}, D.~S. and {R{\'a}cz}, I. and
		  {Rapagnani}, P. and {Razzano}, M. and {Re}, V. and
		  {Regimbau}, T. and {Ricci}, F. and {Robinet}, F. and
		  {Rocchi}, A. and {Rolland}, L. and {Romano}, R. and
		  {Rosi{\'n}ska}, D. and {Ruggi}, P. and {Saracco}, E.},
  title		= "{Advanced Virgo: a second-generation interferometric
		  gravitational wave detector}",
  journal	= {Classical and Quantum Gravity},
  year		= 2015,
  month		= jan,
  volume	= {32},
  number	= {2},
  eid		= {024001},
  pages		= {024001},
  doi		= {10.1088/0264-9381/32/2/024001},
  archiveprefix	= {arXiv},
  eprint	= {1408.3978},
  primaryclass	= {gr-qc},
  adsurl	= {https://ui.adsabs.harvard.edu/abs/2015CQGra..32b4001A}
}

@Article{	  2015cqgra..32g4001l,
  author	= {{LIGO Scientific Collaboration} and {Aasi}, J. and
		  {Abbott}, B.~P. and {Abbott}, R. and {Abbott}, T. and
		  {Abernathy}, M.~R. and {Ackley}, K. and {Adams}, C. and
		  {Adams}, T. and {Addesso}, P. and {Adhikari}, R.~X. and
		  {Adya}, V. and {Affeldt}, C. and {Aggarwal}, N. and
		  {Aguiar}, O.~D. and {Ain}, A. and {Ajith}, P. and {Alemic},
		  A. and {Allen}, B. and {Amariutei}, D. and {Anderson},
		  S.~B. and {Anderson}, W.~G. and {Arai}, K. and {Araya},
		  M.~C. and {Arceneaux}, C. and {Areeda}, J.~S. and {Ashton},
		  G. and {Ast}, S. and {Aston}, S.~M. and {Aufmuth}, P. and
		  {Aulbert}, C. and {Aylott}, B.~E. and {Babak}, S. and
		  {Baker}, P.~T. and {Ballmer}, S.~W. and {Barayoga}, J.~C.
		  and {Barbet}, M. and {Barclay}, S. and {Barish}, B.~C. and
		  {Barker}, D. and {Barr}, B. and {Barsotti}, L. and
		  {Bartlett}, J. and {Barton}, M.~A. and {Bartos}, I. and
		  {Bassiri}, R. and {Batch}, J.~C. and {Baune}, C. and
		  {Behnke}, B. and {Bell}, A.~S. and {Bell}, C. and
		  {Benacquista}, M. and {Bergman}, J. and {Bergmann}, G. and
		  {Berry}, C.~P.~L. and {Betzwieser}, J. and {Bhagwat}, S.
		  and {Bhandare}, R. and {Bilenko}, I.~A. and {Billingsley},
		  G. and {Birch}, J. and {Biscans}, S. and {Biwer}, C. and
		  {Blackburn}, J.~K. and {Blackburn}, L. and {Blair}, C.~D.
		  and {Blair}, D. and {Bock}, O. and {Bodiya}, T.~P. and
		  {Bojtos}, P. and {Bond}, C. and {Bork}, R. and {Born}, M.
		  and {Bose}, Sukanta and {Brady}, P.~R. and {Braginsky},
		  V.~B. and {Brau}, J.~E. and {Bridges}, D.~O. and
		  {Brinkmann}, M. and {Brooks}, A.~F. and {Brown}, D.~A. and
		  {Brown}, D.~D. and {Brown}, N.~M. and {Buchman}, S. and
		  {Buikema}, A. and {Buonanno}, A. and {Cadonati}, L. and
		  {Calder{\'o}n Bustillo}, J. and {Camp}, J.~B. and {Cannon},
		  K.~C. and {Cao}, J. and {Capano}, C.~D. and {Caride}, S.
		  and {Caudill}, S. and {Cavagli{\`a}}, M. and {Cepeda}, C.
		  and {Chakraborty}, R. and {Chalermsongsak}, T. and
		  {Chamberlin}, S.~J. and {Chao}, S. and {Charlton}, P. and
		  {Chen}, Y. and {Cho}, H.~S. and {Cho}, M. and {Chow}, J.~H.
		  and {Christensen}, N. and {Chu}, Q. and {Chung}, S. and
		  {Ciani}, G. and {Clara}, F. and {Clark}, J.~A. and
		  {Collette}, C. and {Cominsky}, L. and {Constancio}, M., Jr.
		  and {Cook}, D. and {Corbitt}, T.~R. and {Cornish}, N. and
		  {Corsi}, A. and {Costa}, C.~A. and {Coughlin}, M.~W. and
		  {Countryman}, S. and {Couvares}, P. and {Coward}, D.~M. and
		  {Cowart}, M.~J. and {Coyne}, D.~C. and {Coyne}, R. and
		  {Craig}, K. and {Creighton}, J.~D.~E. and {Creighton},
		  T.~D. and {Cripe}, J. and {Crowder}, S.~G. and {Cumming},
		  A. and {Cunningham}, L. and {Cutler}, C. and {Dahl}, K. and
		  {Dal Canton}, T. and {Damjanic}, M. and {Danilishin}, S.~L.
		  and {Danzmann}, K. and {Dartez}, L. and {Dave}, I. and
		  {Daveloza}, H. and {Davies}, G.~S. and {Daw}, E.~J. and
		  {DeBra}, D. and {Del Pozzo}, W. and {Denker}, T. and
		  {Dent}, T. and {Dergachev}, V. and {DeRosa}, R.~T. and
		  {DeSalvo}, R. and {Dhurandhar}, S. and
		  {D{\textasciiacute}{\i}az}, M. and {Di Palma}, I. and
		  {Dojcinoski}, G. and {Dominguez}, E. and {Donovan}, F. and
		  {Dooley}, K.~L. and {Doravari}, S. and {Douglas}, R. and
		  {Downes}, T.~P. and {Driggers}, J.~C. and {Du}, Z. and
		  {Dwyer}, S. and {Eberle}, T. and {Edo}, T. and {Edwards},
		  M. and {Edwards}, M. and {Effler}, A. and {Eggenstein}, H.
		  -B. and {Ehrens}, P. and {Eichholz}, J. and {Eikenberry},
		  S.~S. and {Essick}, R. and {Etzel}, T. and {Evans}, M. and
		  {Evans}, T. and {Factourovich}, M. and {Fairhurst}, S. and
		  {Fan}, X. and {Fang}, Q. and {Farr}, B. and {Farr}, W.~M.
		  and {Favata}, M. and {Fays}, M. and {Fehrmann}, H. and
		  {Fejer}, M.~M. and {Feldbaum}, D. and {Ferreira}, E.~C. and
		  {Fisher}, R.~P. and {Frei}, Z. and {Freise}, A. and {Frey},
		  R. and {Fricke}, T.~T. and {Fritschel}, P. and {Frolov},
		  V.~V. and {Fuentes-Tapia}, S. and {Fulda}, P. and {Fyffe},
		  M. and {Gair}, J.~R. and {Gaonkar}, S. and {Gehrels}, N.
		  and {Gergely}, L. {\'A}. and {Giaime}, J.~A. and
		  {Giardina}, K.~D. and {Gleason}, J. and {Goetz}, E. and
		  {Goetz}, R. and {Gondan}, L. and {Gonz{\'a}lez}, G. and
		  {Gordon}, N. and {Gorodetsky}, M.~L. and {Gossan}, S. and
		  {Go{\ss}ler}, S. and {Gr{\"a}f}, C. and {Graff}, P.~B. and
		  {Grant}, A. and {Gras}, S. and {Gray}, C. and {Greenhalgh},
		  R.~J.~S. and {Gretarsson}, A.~M. and {Grote}, H. and
		  {Grunewald}, S. and {Guido}, C.~J. and {Guo}, X. and
		  {Gushwa}, K. and {Gustafson}, E.~K. and {Gustafson}, R. and
		  {Hacker}, J. and {Hall}, E.~D. and {Hammond}, G. and
		  {Hanke}, M. and {Hanks}, J. and {Hanna}, C. and {Hannam},
		  M.~D. and {Hanson}, J. and {Hardwick}, T. and {Harry},
		  G.~M. and {Harry}, I.~W. and {Hart}, M. and {Hartman},
		  M.~T. and {Haster}, C. -J. and {Haughian}, K. and {Hee}, S.
		  and {Heintze}, M. and {Heinzel}, G. and {Hendry}, M. and
		  {Heng}, I.~S. and {Heptonstall}, A.~W. and {Heurs}, M. and
		  {Hewitson}, M. and {Hild}, S. and {Hoak}, D. and {Hodge},
		  K.~A. and {Hollitt}, S.~E. and {Holt}, K. and {Hopkins}, P.
		  and {Hosken}, D.~J. and {Hough}, J. and {Houston}, E. and
		  {Howell}, E.~J. and {Hu}, Y.~M. and {Huerta}, E. and
		  {Hughey}, B. and {Husa}, S. and {Huttner}, S.~H. and
		  {Huynh}, M. and {Huynh-Dinh}, T. and {Idrisy}, A. and
		  {Indik}, N. and {Ingram}, D.~R. and {Inta}, R. and {Islas},
		  G. and {Isler}, J.~C. and {Isogai}, T. and {Iyer}, B.~R.
		  and {Izumi}, K. and {Jacobson}, M. and {Jang}, H. and
		  {Jawahar}, S. and {Ji}, Y. and {Jim{\'e}nez-Forteza}, F.
		  and {Johnson}, W.~W. and {Jones}, D.~I. and {Jones}, R. and
		  {Ju}, L. and {Haris}, K. and {Kalogera}, V. and
		  {Kandhasamy}, S. and {Kang}, G. and {Kanner}, J.~B. and
		  {Katsavounidis}, E. and {Katzman}, W. and {Kaufer}, H. and
		  {Kaufer}, S. and {Kaur}, T. and {Kawabe}, K. and {Kawazoe},
		  F. and {Keiser}, G.~M. and {Keitel}, D. and {Kelley}, D.~B.
		  and {Kells}, W. and {Keppel}, D.~G. and {Key}, J.~S. and
		  {Khalaidovski}, A. and {Khalili}, F.~Y. and {Khazanov},
		  E.~A. and {Kim}, C. and {Kim}, K. and {Kim}, N.~G. and
		  {Kim}, N. and {Kim}, Y. -M. and {King}, E.~J. and {King},
		  P.~J. and {Kinzel}, D.~L. and {Kissel}, J.~S. and
		  {Klimenko}, S. and {Kline}, J. and {Koehlenbeck}, S. and
		  {Kokeyama}, K. and {Kondrashov}, V. and {Korobko}, M. and
		  {Korth}, W.~Z. and {Kozak}, D.~B. and {Kringel}, V. and
		  {Krishnan}, B. and {Krueger}, C. and {Kuehn}, G. and
		  {Kumar}, A. and {Kumar}, P. and {Kuo}, L. and {Landry}, M.
		  and {Lantz}, B. and {Larson}, S. and {Lasky}, P.~D. and
		  {Lazzarini}, A. and {Lazzaro}, C. and {Le}, J. and {Leaci},
		  P. and {Leavey}, S. and {Lebigot}, E.~O. and {Lee}, C.~H.
		  and {Lee}, H.~K. and {Lee}, H.~M. and {Leong}, J.~R. and
		  {Levin}, Y. and {Levine}, B. and {Lewis}, J. and {Li},
		  T.~G.~F. and {Libbrecht}, K. and {Libson}, A. and {Lin},
		  A.~C. and {Littenberg}, T.~B. and {Lockerbie}, N.~A. and
		  {Lockett}, V. and {Logue}, J. and {Lombardi}, A.~L. and
		  {Lormand}, M. and {Lough}, J. and {Lubinski}, M.~J. and
		  {L{\"u}ck}, H. and {Lundgren}, A.~P. and {Lynch}, R. and
		  {Ma}, Y. and {Macarthur}, J. and {MacDonald}, T. and
		  {Machenschalk}, B. and {MacInnis}, M. and {Macleod}, D.~M.
		  and {Maga{\~n}a-Sandoval}, F. and {Magee}, R. and
		  {Mageswaran}, M. and {Maglione}, C. and {Mailand}, K. and
		  {Mandel}, I. and {Mandic}, V. and {Mangano}, V. and
		  {Mansell}, G.~L. and {M{\'a}rka}, S. and {M{\'a}rka}, Z.
		  and {Markosyan}, A. and {Maros}, E. and {Martin}, I.~W. and
		  {Martin}, R.~M. and {Martynov}, D. and {Marx}, J.~N. and
		  {Mason}, K. and {Massinger}, T.~J. and {Matichard}, F. and
		  {Matone}, L. and {Mavalvala}, N. and {Mazumder}, N. and
		  {Mazzolo}, G. and {McCarthy}, R. and {McClelland}, D.~E.
		  and {McCormick}, S. and {McGuire}, S.~C. and {McIntyre}, G.
		  and {McIver}, J. and {McLin}, K. and {McWilliams}, S. and
		  {Meadors}, G.~D. and {Meinders}, M. and {Melatos}, A. and
		  {Mendell}, G. and {Mercer}, R.~A. and {Meshkov}, S. and
		  {Messenger}, C. and {Meyers}, P.~M. and {Miao}, H. and
		  {Middleton}, H. and {Mikhailov}, E.~E. and {Miller}, A. and
		  {Miller}, J. and {Millhouse}, M. and {Ming}, J. and
		  {Mirshekari}, S. and {Mishra}, C. and {Mitra}, S. and
		  {Mitrofanov}, V.~P. and {Mitselmakher}, G. and {Mittleman},
		  R. and {Moe}, B. and {Mohanty}, S.~D. and {Mohapatra},
		  S.~R.~P. and {Moore}, B. and {Moraru}, D. and {Moreno}, G.
		  and {Morriss}, S.~R. and {Mossavi}, K. and {Mow-Lowry},
		  C.~M. and {Mueller}, C.~L. and {Mueller}, G. and
		  {Mukherjee}, S. and {Mullavey}, A. and {Munch}, J. and
		  {Murphy}, D. and {Murray}, P.~G. and {Mytidis}, A. and
		  {Nash}, T. and {Nayak}, R.~K. and {Necula}, V. and
		  {Nedkova}, K. and {Newton}, G. and {Nguyen}, T. and
		  {Nielsen}, A.~B. and {Nissanke}, S. and {Nitz}, A.~H. and
		  {Nolting}, D. and {Normandin}, M.~E.~N. and {Nuttall},
		  L.~K. and {Ochsner}, E. and {O'Dell}, J. and {Oelker}, E.
		  and {Ogin}, G.~H. and {Oh}, J.~J. and {Oh}, S.~H. and
		  {Ohme}, F. and {Oppermann}, P. and {Oram}, R. and
		  {O'Reilly}, B. and {Ortega}, W. and {O'Shaughnessy}, R. and
		  {Osthelder}, C. and {Ott}, C.~D. and {Ottaway}, D.~J. and
		  {Ottens}, R.~S. and {Overmier}, H. and {Owen}, B.~J. and
		  {Padilla}, C. and {Pai}, A. and {Pai}, S. and {Palashov},
		  O. and {Pal-Singh}, A. and {Pan}, H. and {Pankow}, C. and
		  {Pannarale}, F. and {Pant}, B.~C. and {Papa}, M.~A. and
		  {Paris}, H. and {Patrick}, Z. and {Pedraza}, M. and
		  {Pekowsky}, L. and {Pele}, A. and {Penn}, S. and {Perreca},
		  A. and {Phelps}, M. and {Pierro}, V. and {Pinto}, I.~M. and
		  {Pitkin}, M. and {Poeld}, J. and {Post}, A. and
		  {Poteomkin}, A. and {Powell}, J. and {Prasad}, J. and
		  {Predoi}, V. and {Premachandra}, S. and {Prestegard}, T.
		  and {Price}, L.~R. and {Principe}, M. and {Privitera}, S.
		  and {Prix}, R. and {Prokhorov}, L. and {Puncken}, O. and
		  {P{\"u}rrer}, M. and {Qin}, J. and {Quetschke}, V. and
		  {Quintero}, E. and {Quiroga}, G. and {Quitzow-James}, R.
		  and {Raab}, F.~J. and {Rabeling}, D.~S. and {Radkins}, H.
		  and {Raffai}, P. and {Raja}, S. and {Rajalakshmi}, G. and
		  {Rakhmanov}, M. and {Ramirez}, K. and {Raymond}, V. and
		  {Reed}, C.~M. and {Reid}, S. and {Reitze}, D.~H. and
		  {Reula}, O. and {Riles}, K. and {Robertson}, N.~A. and
		  {Robie}, R. and {Rollins}, J.~G. and {Roma}, V. and
		  {Romano}, J.~D. and {Romanov}, G. and {Romie}, J.~H. and
		  {Rowan}, S. and {R{\"u}diger}, A. and {Ryan}, K. and
		  {Sachdev}, S. and {Sadecki}, T. and {Sadeghian}, L. and
		  {Saleem}, M. and {Salemi}, F. and {Sammut}, L. and
		  {Sandberg}, V. and {Sanders}, J.~R. and {Sannibale}, V. and
		  {Santiago-Prieto}, I. and {Sathyaprakash}, B.~S. and
		  {Saulson}, P.~R. and {Savage}, R. and {Sawadsky}, A. and
		  {Scheuer}, J. and {Schilling}, R. and {Schmidt}, P. and
		  {Schnabel}, R. and {Schofield}, R.~M.~S. and {Schreiber},
		  E. and {Schuette}, D. and {Schutz}, B.~F. and {Scott}, J.
		  and {Scott}, S.~M. and {Sellers}, D. and {Sengupta}, A.~S.
		  and {Sergeev}, A. and {Serna}, G. and {Sevigny}, A. and
		  {Shaddock}, D.~A. and {Shahriar}, M.~S. and {Shaltev}, M.
		  and {Shao}, Z. and {Shapiro}, B. and {Shawhan}, P. and
		  {Shoemaker}, D.~H. and {Sidery}, T.~L. and {Siemens}, X.
		  and {Sigg}, D. and {Silva}, A.~D. and {Simakov}, D. and
		  {Singer}, A. and {Singer}, L. and {Singh}, R. and {Sintes},
		  A.~M. and {Slagmolen}, B.~J.~J. and {Smith}, J.~R. and
		  {Smith}, M.~R. and {Smith}, R.~J.~E. and {Smith-Lefebvre},
		  N.~D. and {Son}, E.~J. and {Sorazu}, B. and {Souradeep}, T.
		  and {Staley}, A. and {Stebbins}, J. and {Steinke}, M. and
		  {Steinlechner}, J. and {Steinlechner}, S. and {Steinmeyer},
		  D. and {Stephens}, B.~C. and {Steplewski}, S. and
		  {Stevenson}, S. and {Stone}, R. and {Strain}, K.~A. and
		  {Strigin}, S. and {Sturani}, R. and {Stuver}, A.~L. and
		  {Summerscales}, T.~Z. and {Sutton}, P.~J. and
		  {Szczepanczyk}, M. and {Szeifert}, G. and {Talukder}, D.
		  and {Tanner}, D.~B. and {T{\'a}pai}, M. and {Tarabrin},
		  S.~P. and {Taracchini}, A. and {Taylor}, R. and {Tellez},
		  G. and {Theeg}, T. and {Thirugnanasambandam}, M.~P. and
		  {Thomas}, M. and {Thomas}, P. and {Thorne}, K.~A. and
		  {Thorne}, K.~S. and {Thrane}, E. and {Tiwari}, V. and
		  {Tomlinson}, C. and {Torres}, C.~V. and {Torrie}, C.~I. and
		  {Traylor}, G. and {Tse}, M. and {Tshilumba}, D. and
		  {Ugolini}, D. and {Unnikrishnan}, C.~S. and {Urban}, A.~L.
		  and {Usman}, S.~A. and {Vahlbruch}, H. and {Vajente}, G.
		  and {Valdes}, G. and {Vallisneri}, M. and {van Veggel},
		  A.~A. and {Vass}, S. and {Vaulin}, R. and {Vecchio}, A. and
		  {Veitch}, J. and {Veitch}, P.~J. and {Venkateswara}, K. and
		  {Vincent-Finley}, R. and {Vitale}, S. and {Vo}, T. and
		  {Vorvick}, C. and {Vousden}, W.~D. and {Vyatchanin}, S.~P.
		  and {Wade}, A.~R. and {Wade}, L. and {Wade}, M. and
		  {Walker}, M. and {Wallace}, L. and {Walsh}, S. and {Wang},
		  H. and {Wang}, M. and {Wang}, X. and {Ward}, R.~L. and
		  {Warner}, J. and {Was}, M. and {Weaver}, B. and {Weinert},
		  M. and {Weinstein}, A.~J. and {Weiss}, R. and {Welborn}, T.
		  and {Wen}, L. and {Wessels}, P. and {Westphal}, T. and
		  {Wette}, K. and {Whelan}, J.~T. and {Whitcomb}, S.~E. and
		  {White}, D.~J. and {Whiting}, B.~F. and {Wilkinson}, C. and
		  {Williams}, L. and {Williams}, R. and {Williamson}, A.~R.
		  and {Willis}, J.~L. and {Willke}, B. and {Wimmer}, M. and
		  {Winkler}, W. and {Wipf}, C.~C. and {Wittel}, H. and
		  {Woan}, G. and {Worden}, J. and {Xie}, S. and {Yablon}, J.
		  and {Yakushin}, I. and {Yam}, W. and {Yamamoto}, H. and
		  {Yancey}, C.~C. and {Yang}, Q. and {Zanolin}, M. and
		  {Zhang}, Fan and {Zhang}, L. and {Zhang}, M. and {Zhang},
		  Y. and {Zhao}, C. and {Zhou}, M. and {Zhu}, X.~J. and
		  {Zucker}, M.~E. and {Zuraw}, S. and {Zweizig}, J.},
  title		= {{Advanced LIGO}},
  journal	= {Classical and Quantum Gravity},
  year		= 2015,
  month		= apr,
  volume	= {32},
  number	= {7},
  eid		= {074001},
  pages		= {074001},
  doi		= {10.1088/0264-9381/32/7/074001},
  archiveprefix	= {arXiv},
  eprint	= {1411.4547},
  primaryclass	= {gr-qc},
  adsurl	= {https://ui.adsabs.harvard.edu/abs/2015CQGra..32g4001L}
}

@Article{	  2015phrvd..91h4011a,
  author	= {{Arvanitaki}, Asimina and {Baryakhtar}, Masha and {Huang},
		  Xinlu},
  title		= "{Discovering the QCD axion with black holes and
		  gravitational waves}",
  journal	= {\prd},
  year		= 2015,
  month		= apr,
  volume	= {91},
  number	= {8},
  eid		= {084011},
  pages		= {084011},
  doi		= {10.1103/PhysRevD.91.084011},
  archiveprefix	= {arXiv},
  eprint	= {1411.2263},
  primaryclass	= {hep-ph},
  adsurl	= {https://ui.adsabs.harvard.edu/abs/2015PhRvD..91h4011A}
}

@Article{	  2016phrvl.116f1102a,
  author	= {{Abbott}, B.~P. and {Abbott}, R. and {Abbott}, T.~D. and
		  {Abernathy}, M.~R. and {Acernese}, F. and {Ackley}, K. and
		  {Adams}, C. and {Adams}, T. and {Addesso}, P. and
		  {Adhikari}, R.~X. and {Adya}, V.~B. and {Affeldt}, C. and
		  {Agathos}, M. and {Agatsuma}, K. and {Aggarwal}, N. and
		  {Aguiar}, O.~D. and {Aiello}, L. and {Ain}, A. and {Ajith},
		  P. and {Allen}, B. and {Allocca}, A. and {Altin}, P.~A. and
		  {Anderson}, S.~B. and {Anderson}, W.~G. and {Arai}, K. and
		  {Arain}, M.~A. and {Araya}, M.~C. and {Arceneaux}, C.~C.
		  and {Areeda}, J.~S. and {Arnaud}, N. and {Arun}, K.~G. and
		  {Ascenzi}, S. and {Ashton}, G. and {Ast}, M. and {Aston},
		  S.~M. and {Astone}, P. and {Aufmuth}, P. and {Aulbert}, C.
		  and {Babak}, S. and {Bacon}, P. and {Bader}, M.~K.~M. and
		  {Baker}, P.~T. and {Baldaccini}, F. and {Ballardin}, G. and
		  {Ballmer}, S.~W. and {Barayoga}, J.~C. and {Barclay}, S.~E.
		  and {Barish}, B.~C. and {Barker}, D. and {Barone}, F. and
		  {Barr}, B. and {Barsotti}, L. and {Barsuglia}, M. and
		  {Barta}, D. and {Bartlett}, J. and {Barton}, M.~A. and
		  {Bartos}, I. and {Bassiri}, R. and {Basti}, A. and {Batch},
		  J.~C. and {Baune}, C. and {Bavigadda}, V. and {Bazzan}, M.
		  and {Behnke}, B. and {Bejger}, M. and {Belczynski}, C. and
		  {Bell}, A.~S. and {Bell}, C.~J. and {Berger}, B.~K. and
		  {Bergman}, J. and {Bergmann}, G. and {Berry}, C.~P.~L. and
		  {Bersanetti}, D. and {Bertolini}, A. and {Betzwieser}, J.
		  and {Bhagwat}, S. and {Bhandare}, R. and {Bilenko}, I.~A.
		  and {Billingsley}, G. and {Birch}, J. and {Birney}, R. and
		  {Birnholtz}, O. and {Biscans}, S. and {Bisht}, A. and
		  {Bitossi}, M. and {Biwer}, C. and {Bizouard}, M.~A. and
		  {Blackburn}, J.~K. and {Blair}, C.~D. and {Blair}, D.~G.
		  and {Blair}, R.~M. and {Bloemen}, S. and {Bock}, O. and
		  {Bodiya}, T.~P. and {Boer}, M. and {Bogaert}, G. and
		  {Bogan}, C. and {Bohe}, A. and {Bojtos}, P. and {Bond}, C.
		  and {Bondu}, F. and {Bonnand}, R. and {Boom}, B.~A. and
		  {Bork}, R. and {Boschi}, V. and {Bose}, S. and
		  {Bouffanais}, Y. and {Bozzi}, A. and {Bradaschia}, C. and
		  {Brady}, P.~R. and {Braginsky}, V.~B. and {Branchesi}, M.
		  and {Brau}, J.~E. and {Briant}, T. and {Brillet}, A. and
		  {Brinkmann}, M. and {Brisson}, V. and {Brockill}, P. and
		  {Brooks}, A.~F. and {Brown}, D.~A. and {Brown}, D.~D. and
		  {Brown}, N.~M. and {Buchanan}, C.~C. and {Buikema}, A. and
		  {Bulik}, T. and {Bulten}, H.~J. and {Buonanno}, A. and
		  {Buskulic}, D. and {Buy}, C. and {Byer}, R.~L. and
		  {Cabero}, M. and {Cadonati}, L. and {Cagnoli}, G. and
		  {Cahillane}, C. and {Bustillo}, J. Calder{\'o}n and
		  {Callister}, T. and {Calloni}, E. and {Camp}, J.~B. and
		  {Cannon}, K.~C. and {Cao}, J. and {Capano}, C.~D. and
		  {Capocasa}, E. and {Carbognani}, F. and {Caride}, S. and
		  {Casanueva Diaz}, J. and {Casentini}, C. and {Caudill}, S.
		  and {Cavagli{\`a}}, M. and {Cavalier}, F. and {Cavalieri},
		  R. and {Cella}, G. and {Cepeda}, C.~B. and {Baiardi}, L.
		  Cerboni and {Cerretani}, G. and {Cesarini}, E. and
		  {Chakraborty}, R. and {Chalermsongsak}, T. and
		  {Chamberlin}, S.~J. and {Chan}, M. and {Chao}, S. and
		  {Charlton}, P. and {Chassande-Mottin}, E. and {Chen}, H.~Y.
		  and {Chen}, Y. and {Cheng}, C. and {Chincarini}, A. and
		  {Chiummo}, A. and {Cho}, H.~S. and {Cho}, M. and {Chow},
		  J.~H. and {Christensen}, N. and {Chu}, Q. and {Chua}, S.
		  and {Chung}, S. and {Ciani}, G. and {Clara}, F. and
		  {Clark}, J.~A. and {Cleva}, F. and {Coccia}, E. and
		  {Cohadon}, P. -F. and {Colla}, A. and {Collette}, C.~G. and
		  {Cominsky}, L. and {Constancio}, M. and {Conte}, A. and
		  {Conti}, L. and {Cook}, D. and {Corbitt}, T.~R. and
		  {Cornish}, N. and {Corsi}, A. and {Cortese}, S. and
		  {Costa}, C.~A. and {Coughlin}, M.~W. and {Coughlin}, S.~B.
		  and {Coulon}, J. -P. and {Countryman}, S.~T. and
		  {Couvares}, P. and {Cowan}, E.~E. and {Coward}, D.~M. and
		  {Cowart}, M.~J. and {Coyne}, D.~C. and {Coyne}, R. and
		  {Craig}, K. and {Creighton}, J.~D.~E. and {Creighton},
		  T.~D. and {Cripe}, J. and {Crowder}, S.~G. and {Cruise},
		  A.~M. and {Cumming}, A. and {Cunningham}, L. and {Cuoco},
		  E. and {Dal Canton}, T. and {Danilishin}, S.~L. and
		  {D'Antonio}, S. and {Danzmann}, K. and {Darman}, N.~S. and
		  {Da Silva Costa}, C.~F. and {Dattilo}, V. and {Dave}, I.
		  and {Daveloza}, H.~P. and {Davier}, M. and {Davies}, G.~S.
		  and {Daw}, E.~J. and {Day}, R. and {De}, S. and {DeBra}, D.
		  and {Debreczeni}, G. and {Degallaix}, J. and {De
		  Laurentis}, M. and {Del{\'e}glise}, S. and {Del Pozzo}, W.
		  and {Denker}, T. and {Dent}, T. and {Dereli}, H. and
		  {Dergachev}, V. and {DeRosa}, R.~T. and {De Rosa}, R. and
		  {DeSalvo}, R. and {Dhurandhar}, S. and {D{\'\i}az}, M.~C.
		  and {Di Fiore}, L. and {Di Giovanni}, M. and {Di Lieto}, A.
		  and {Di Pace}, S. and {Di Palma}, I. and {Di Virgilio}, A.
		  and {Dojcinoski}, G. and {Dolique}, V. and {Donovan}, F.
		  and {Dooley}, K.~L. and {Doravari}, S. and {Douglas}, R.
		  and {Downes}, T.~P. and {Drago}, M. and {Drever}, R.~W.~P.
		  and {Driggers}, J.~C. and {Du}, Z. and {Ducrot}, M. and
		  {Dwyer}, S.~E. and {Edo}, T.~B. and {Edwards}, M.~C. and
		  {Effler}, A. and {Eggenstein}, H. -B. and {Ehrens}, P. and
		  {Eichholz}, J. and {Eikenberry}, S.~S. and {Engels}, W. and
		  {Essick}, R.~C. and {Etzel}, T. and {Evans}, M. and
		  {Evans}, T.~M. and {Everett}, R. and {Factourovich}, M. and
		  {Fafone}, V. and {Fair}, H. and {Fairhurst}, S. and {Fan},
		  X. and {Fang}, Q. and {Farinon}, S. and {Farr}, B. and
		  {Farr}, W.~M. and {Favata}, M. and {Fays}, M. and
		  {Fehrmann}, H. and {Fejer}, M.~M. and {Feldbaum}, D. and
		  {Ferrante}, I. and {Ferreira}, E.~C. and {Ferrini}, F. and
		  {Fidecaro}, F. and {Finn}, L.~S. and {Fiori}, I. and
		  {Fiorucci}, D. and {Fisher}, R.~P. and {Flaminio}, R. and
		  {Fletcher}, M. and {Fong}, H. and {Fournier}, J. -D. and
		  {Franco}, S. and {Frasca}, S. and {Frasconi}, F. and
		  {Frede}, M. and {Frei}, Z. and {Freise}, A. and {Frey}, R.
		  and {Frey}, V. and {Fricke}, T.~T. and {Fritschel}, P. and
		  {Frolov}, V.~V. and {Fulda}, P. and {Fyffe}, M. and
		  {Gabbard}, H.~A.~G. and {Gair}, J.~R. and {Gammaitoni}, L.
		  and {Gaonkar}, S.~G. and {Garufi}, F. and {Gatto}, A. and
		  {Gaur}, G. and {Gehrels}, N. and {Gemme}, G. and {Gendre},
		  B. and {Genin}, E. and {Gennai}, A. and {George}, J. and
		  {Gergely}, L. and {Germain}, V. and {Ghosh}, Abhirup and
		  {Ghosh}, Archisman and {Ghosh}, S. and {Giaime}, J.~A. and
		  {Giardina}, K.~D. and {Giazotto}, A. and {Gill}, K. and
		  {Glaefke}, A. and {Gleason}, J.~R. and {Goetz}, E. and
		  {Goetz}, R. and {Gondan}, L. and {Gonz{\'a}lez}, G. and
		  {Castro}, J.~M. Gonzalez and {Gopakumar}, A. and {Gordon},
		  N.~A. and {Gorodetsky}, M.~L. and {Gossan}, S.~E. and
		  {Gosselin}, M. and {Gouaty}, R. and {Graef}, C. and
		  {Graff}, P.~B. and {Granata}, M. and {Grant}, A. and
		  {Gras}, S. and {Gray}, C. and {Greco}, G. and {Green},
		  A.~C. and {Greenhalgh}, R.~J.~S. and {Groot}, P. and
		  {Grote}, H. and {Grunewald}, S. and {Guidi}, G.~M. and
		  {Guo}, X. and {Gupta}, A. and {Gupta}, M.~K. and {Gushwa},
		  K.~E. and {Gustafson}, E.~K. and {Gustafson}, R. and
		  {Hacker}, J.~J. and {Hall}, B.~R. and {Hall}, E.~D. and
		  {Hammond}, G. and {Haney}, M. and {Hanke}, M.~M. and
		  {Hanks}, J. and {Hanna}, C. and {Hannam}, M.~D. and
		  {Hanson}, J. and {Hardwick}, T. and {Harms}, J. and
		  {Harry}, G.~M. and {Harry}, I.~W. and {Hart}, M.~J. and
		  {Hartman}, M.~T. and {Haster}, C. -J. and {Haughian}, K.
		  and {Healy}, J. and {Heefner}, J. and {Heidmann}, A. and
		  {Heintze}, M.~C. and {Heinzel}, G. and {Heitmann}, H. and
		  {Hello}, P. and {Hemming}, G. and {Hendry}, M. and {Heng},
		  I.~S. and {Hennig}, J. and {Heptonstall}, A.~W. and
		  {Heurs}, M. and {Hild}, S. and {Hoak}, D. and {Hodge},
		  K.~A. and {Hofman}, D. and {Hollitt}, S.~E. and {Holt}, K.
		  and {Holz}, D.~E. and {Hopkins}, P. and {Hosken}, D.~J. and
		  {Hough}, J. and {Houston}, E.~A. and {Howell}, E.~J. and
		  {Hu}, Y.~M. and {Huang}, S. and {Huerta}, E.~A. and {Huet},
		  D. and {Hughey}, B. and {Husa}, S. and {Huttner}, S.~H. and
		  {Huynh-Dinh}, T. and {Idrisy}, A. and {Indik}, N. and
		  {Ingram}, D.~R. and {Inta}, R. and {Isa}, H.~N. and {Isac},
		  J. -M. and {Isi}, M. and {Islas}, G. and {Isogai}, T. and
		  {Iyer}, B.~R. and {Izumi}, K. and {Jacobson}, M.~B. and
		  {Jacqmin}, T. and {Jang}, H. and {Jani}, K. and
		  {Jaranowski}, P. and {Jawahar}, S. and
		  {Jim{\'e}nez-Forteza}, F. and {Johnson}, W.~W. and
		  {Johnson-McDaniel}, N.~K. and {Jones}, D.~I. and {Jones},
		  R. and {Jonker}, R.~J.~G. and {Ju}, L. and {Haris}, K. and
		  {Kalaghatgi}, C.~V. and {Kalogera}, V. and {Kandhasamy}, S.
		  and {Kang}, G. and {Kanner}, J.~B. and {Karki}, S. and
		  {Kasprzack}, M. and {Katsavounidis}, E. and {Katzman}, W.
		  and {Kaufer}, S. and {Kaur}, T. and {Kawabe}, K. and
		  {Kawazoe}, F. and {K{\'e}f{\'e}lian}, F. and {Kehl}, M.~S.
		  and {Keitel}, D. and {Kelley}, D.~B. and {Kells}, W. and
		  {Kennedy}, R. and {Keppel}, D.~G. and {Key}, J.~S. and
		  {Khalaidovski}, A. and {Khalili}, F.~Y. and {Khan}, I. and
		  {Khan}, S. and {Khan}, Z. and {Khazanov}, E.~A. and
		  {Kijbunchoo}, N. and {Kim}, C. and {Kim}, J. and {Kim}, K.
		  and {Kim}, Nam-Gyu and {Kim}, Namjun and {Kim}, Y. -M. and
		  {King}, E.~J. and {King}, P.~J. and {Kinzel}, D.~L. and
		  {Kissel}, J.~S. and {Kleybolte}, L. and {Klimenko}, S. and
		  {Koehlenbeck}, S.~M. and {Kokeyama}, K. and {Koley}, S. and
		  {Kondrashov}, V. and {Kontos}, A. and {Koranda}, S. and
		  {Korobko}, M. and {Korth}, W.~Z. and {Kowalska}, I. and
		  {Kozak}, D.~B. and {Kringel}, V. and {Krishnan}, B. and
		  {Kr{\'o}lak}, A. and {Krueger}, C. and {Kuehn}, G. and
		  {Kumar}, P. and {Kumar}, R. and {Kuo}, L. and {Kutynia}, A.
		  and {Kwee}, P. and {Lackey}, B.~D. and {Landry}, M. and
		  {Lange}, J. and {Lantz}, B. and {Lasky}, P.~D. and
		  {Lazzarini}, A. and {Lazzaro}, C. and {Leaci}, P. and
		  {Leavey}, S. and {Lebigot}, E.~O. and {Lee}, C.~H. and
		  {Lee}, H.~K. and {Lee}, H.~M. and {Lee}, K. and {Lenon}, A.
		  and {Leonardi}, M. and {Leong}, J.~R. and {Leroy}, N. and
		  {Letendre}, N. and {Levin}, Y. and {Levine}, B.~M. and
		  {Li}, T.~G.~F. and {Libson}, A. and {Littenberg}, T.~B. and
		  {Lockerbie}, N.~A. and {Logue}, J. and {Lombardi}, A.~L.
		  and {London}, L.~T. and {Lord}, J.~E. and {Lorenzini}, M.
		  and {Loriette}, V. and {Lormand}, M. and {Losurdo}, G. and
		  {Lough}, J.~D. and {Lousto}, C.~O. and {Lovelace}, G. and
		  {L{\"u}ck}, H. and {Lundgren}, A.~P. and {Luo}, J. and
		  {Lynch}, R. and {Ma}, Y. and {MacDonald}, T. and
		  {Machenschalk}, B. and {MacInnis}, M. and {Macleod}, D.~M.
		  and {Maga{\~n}a-Sandoval}, F. and {Magee}, R.~M. and
		  {Mageswaran}, M. and {Majorana}, E. and {Maksimovic}, I.
		  and {Malvezzi}, V. and {Man}, N. and {Mandel}, I. and
		  {Mandic}, V. and {Mangano}, V. and {Mansell}, G.~L. and
		  {Manske}, M. and {Mantovani}, M. and {Marchesoni}, F. and
		  {Marion}, F. and {M{\'a}rka}, S. and {M{\'a}rka}, Z. and
		  {Markosyan}, A.~S. and {Maros}, E. and {Martelli}, F. and
		  {Martellini}, L. and {Martin}, I.~W. and {Martin}, R.~M.
		  and {Martynov}, D.~V. and {Marx}, J.~N. and {Mason}, K. and
		  {Masserot}, A. and {Massinger}, T.~J. and {Masso-Reid}, M.
		  and {Matichard}, F. and {Matone}, L. and {Mavalvala}, N.
		  and {Mazumder}, N. and {Mazzolo}, G. and {McCarthy}, R. and
		  {McClelland}, D.~E. and {McCormick}, S. and {McGuire},
		  S.~C. and {McIntyre}, G. and {McIver}, J. and {McManus},
		  D.~J. and {McWilliams}, S.~T. and {Meacher}, D. and
		  {Meadors}, G.~D. and {Meidam}, J. and {Melatos}, A. and
		  {Mendell}, G. and {Mendoza-Gandara}, D. and {Mercer}, R.~A.
		  and {Merilh}, E. and {Merzougui}, M. and {Meshkov}, S. and
		  {Messenger}, C. and {Messick}, C. and {Meyers}, P.~M. and
		  {Mezzani}, F. and {Miao}, H. and {Michel}, C. and
		  {Middleton}, H. and {Mikhailov}, E.~E. and {Milano}, L. and
		  {Miller}, J. and {Millhouse}, M. and {Minenkov}, Y. and
		  {Ming}, J. and {Mirshekari}, S. and {Mishra}, C. and
		  {Mitra}, S. and {Mitrofanov}, V.~P. and {Mitselmakher}, G.
		  and {Mittleman}, R. and {Moggi}, A. and {Mohan}, M. and
		  {Mohapatra}, S.~R.~P. and {Montani}, M. and {Moore}, B.~C.
		  and {Moore}, C.~J. and {Moraru}, D. and {Moreno}, G. and
		  {Morriss}, S.~R. and {Mossavi}, K. and {Mours}, B. and
		  {Mow-Lowry}, C.~M. and {Mueller}, C.~L. and {Mueller}, G.
		  and {Muir}, A.~W. and {Mukherjee}, Arunava and {Mukherjee},
		  D. and {Mukherjee}, S. and {Mukund}, N. and {Mullavey}, A.
		  and {Munch}, J. and {Murphy}, D.~J. and {Murray}, P.~G. and
		  {Mytidis}, A. and {Nardecchia}, I. and {Naticchioni}, L.
		  and {Nayak}, R.~K. and {Necula}, V. and {Nedkova}, K. and
		  {Nelemans}, G. and {Neri}, M. and {Neunzert}, A. and
		  {Newton}, G. and {Nguyen}, T.~T. and {Nielsen}, A.~B. and
		  {Nissanke}, S. and {Nitz}, A. and {Nocera}, F. and
		  {Nolting}, D. and {Normandin}, M.~E.~N. and {Nuttall},
		  L.~K. and {Oberling}, J. and {Ochsner}, E. and {O'Dell}, J.
		  and {Oelker}, E. and {Ogin}, G.~H. and {Oh}, J.~J. and
		  {Oh}, S.~H. and {Ohme}, F. and {Oliver}, M. and
		  {Oppermann}, P. and {Oram}, Richard J. and {O'Reilly}, B.
		  and {O'Shaughnessy}, R. and {Ott}, C.~D. and {Ottaway},
		  D.~J. and {Ottens}, R.~S. and {Overmier}, H. and {Owen},
		  B.~J. and {Pai}, A. and {Pai}, S.~A. and {Palamos}, J.~R.
		  and {Palashov}, O. and {Palomba}, C. and {Pal-Singh}, A.
		  and {Pan}, H. and {Pan}, Y. and {Pankow}, C. and
		  {Pannarale}, F. and {Pant}, B.~C. and {Paoletti}, F. and
		  {Paoli}, A. and {Papa}, M.~A. and {Paris}, H.~R. and
		  {Parker}, W. and {Pascucci}, D. and {Pasqualetti}, A. and
		  {Passaquieti}, R. and {Passuello}, D. and {Patricelli}, B.
		  and {Patrick}, Z. and {Pearlstone}, B.~L. and {Pedraza}, M.
		  and {Pedurand}, R. and {Pekowsky}, L. and {Pele}, A. and
		  {Penn}, S. and {Perreca}, A. and {Pfeiffer}, H.~P. and
		  {Phelps}, M. and {Piccinni}, O. and {Pichot}, M. and
		  {Pickenpack}, M. and {Piergiovanni}, F. and {Pierro}, V.
		  and {Pillant}, G. and {Pinard}, L. and {Pinto}, I.~M. and
		  {Pitkin}, M. and {Poeld}, J.~H. and {Poggiani}, R. and
		  {Popolizio}, P. and {Post}, A. and {Powell}, J. and
		  {Prasad}, J. and {Predoi}, V. and {Premachandra}, S.~S. and
		  {Prestegard}, T. and {Price}, L.~R. and {Prijatelj}, M. and
		  {Principe}, M. and {Privitera}, S. and {Prix}, R. and
		  {Prodi}, G.~A. and {Prokhorov}, L. and {Puncken}, O. and
		  {Punturo}, M. and {Puppo}, P. and {P{\"u}rrer}, M. and
		  {Qi}, H. and {Qin}, J. and {Quetschke}, V. and {Quintero},
		  E.~A. and {Quitzow-James}, R. and {Raab}, F.~J. and
		  {Rabeling}, D.~S. and {Radkins}, H. and {Raffai}, P. and
		  {Raja}, S. and {Rakhmanov}, M. and {Ramet}, C.~R. and
		  {Rapagnani}, P. and {Raymond}, V. and {Razzano}, M. and
		  {Re}, V. and {Read}, J. and {Reed}, C.~M. and {Regimbau},
		  T. and {Rei}, L. and {Reid}, S. and {Reitze}, D.~H. and
		  {Rew}, H. and {Reyes}, S.~D. and {Ricci}, F. and {Riles},
		  K. and {Robertson}, N.~A. and {Robie}, R. and {Robinet}, F.
		  and {Rocchi}, A. and {Rolland}, L. and {Rollins}, J.~G. and
		  {Roma}, V.~J. and {Romano}, J.~D. and {Romano}, R. and
		  {Romanov}, G. and {Romie}, J.~H. and {Rosi{\'n}ska}, D. and
		  {Rowan}, S. and {R{\"u}diger}, A. and {Ruggi}, P. and
		  {Ryan}, K. and {Sachdev}, S. and {Sadecki}, T. and
		  {Sadeghian}, L. and {Salconi}, L. and {Saleem}, M. and
		  {Salemi}, F. and {Samajdar}, A. and {Sammut}, L. and
		  {Sampson}, L.~M. and {Sanchez}, E.~J. and {Sandberg}, V.
		  and {Sandeen}, B. and {Sanders}, G.~H. and {Sanders}, J.~R.
		  and {Sassolas}, B. and {Sathyaprakash}, B.~S. and
		  {Saulson}, P.~R. and {Sauter}, O. and {Savage}, R.~L. and
		  {Sawadsky}, A. and {Schale}, P. and {Schilling}, R. and
		  {Schmidt}, J. and {Schmidt}, P. and {Schnabel}, R. and
		  {Schofield}, R.~M.~S. and {Sch{\"o}nbeck}, A. and
		  {Schreiber}, E. and {Schuette}, D. and {Schutz}, B.~F. and
		  {Scott}, J. and {Scott}, S.~M. and {Sellers}, D. and
		  {Sengupta}, A.~S. and {Sentenac}, D. and {Sequino}, V. and
		  {Sergeev}, A. and {Serna}, G. and {Setyawati}, Y. and
		  {Sevigny}, A. and {Shaddock}, D.~A. and {Shaffer}, T. and
		  {Shah}, S. and {Shahriar}, M.~S. and {Shaltev}, M. and
		  {Shao}, Z. and {Shapiro}, B. and {Shawhan}, P. and
		  {Sheperd}, A. and {Shoemaker}, D.~H. and {Shoemaker}, D.~M.
		  and {Siellez}, K. and {Siemens}, X. and {Sigg}, D. and
		  {Silva}, A.~D. and {Simakov}, D. and {Singer}, A. and
		  {Singer}, L.~P. and {Singh}, A. and {Singh}, R. and
		  {Singhal}, A. and {Sintes}, A.~M. and {Slagmolen}, B.~J.~J.
		  and {Smith}, J.~R. and {Smith}, M.~R. and {Smith}, N.~D.
		  and {Smith}, R.~J.~E. and {Son}, E.~J. and {Sorazu}, B. and
		  {Sorrentino}, F. and {Souradeep}, T. and {Srivastava},
		  A.~K. and {Staley}, A. and {Steinke}, M. and
		  {Steinlechner}, J. and {Steinlechner}, S. and {Steinmeyer},
		  D. and {Stephens}, B.~C. and {Stevenson}, S.~P. and
		  {Stone}, R. and {Strain}, K.~A. and {Straniero}, N. and
		  {Stratta}, G. and {Strauss}, N.~A. and {Strigin}, S. and
		  {Sturani}, R. and {Stuver}, A.~L. and {Summerscales}, T.~Z.
		  and {Sun}, L. and {Sutton}, P.~J. and {Swinkels}, B.~L. and
		  {Szczepa{\'n}czyk}, M.~J. and {Tacca}, M. and {Talukder},
		  D. and {Tanner}, D.~B. and {T{\'a}pai}, M. and {Tarabrin},
		  S.~P. and {Taracchini}, A. and {Taylor}, R. and {Theeg}, T.
		  and {Thirugnanasambandam}, M.~P. and {Thomas}, E.~G. and
		  {Thomas}, M. and {Thomas}, P. and {Thorne}, K.~A. and
		  {Thorne}, K.~S. and {Thrane}, E. and {Tiwari}, S. and
		  {Tiwari}, V. and {Tokmakov}, K.~V. and {Tomlinson}, C. and
		  {Tonelli}, M. and {Torres}, C.~V. and {Torrie}, C.~I. and
		  {T{\"o}yr{\"a}}, D. and {Travasso}, F. and {Traylor}, G.
		  and {Trifir{\`o}}, D. and {Tringali}, M.~C. and {Trozzo},
		  L. and {Tse}, M. and {Turconi}, M. and {Tuyenbayev}, D. and
		  {Ugolini}, D. and {Unnikrishnan}, C.~S. and {Urban}, A.~L.
		  and {Usman}, S.~A. and {Vahlbruch}, H. and {Vajente}, G.
		  and {Valdes}, G. and {Vallisneri}, M. and {van Bakel}, N.
		  and {van Beuzekom}, M. and {van den Brand}, J.~F.~J. and
		  {Van Den Broeck}, C. and {Vander-Hyde}, D.~C. and {van der
		  Schaaf}, L. and {van Heijningen}, J.~V. and {van Veggel},
		  A.~A. and {Vardaro}, M. and {Vass}, S. and {Vas{\'u}th}, M.
		  and {Vaulin}, R. and {Vecchio}, A. and {Vedovato}, G. and
		  {Veitch}, J. and {Veitch}, P.~J. and {Venkateswara}, K. and
		  {Verkindt}, D. and {Vetrano}, F. and {Vicer{\'e}}, A. and
		  {Vinciguerra}, S. and {Vine}, D.~J. and {Vinet}, J. -Y. and
		  {Vitale}, S. and {Vo}, T. and {Vocca}, H. and {Vorvick}, C.
		  and {Voss}, D. and {Vousden}, W.~D. and {Vyatchanin}, S.~P.
		  and {Wade}, A.~R. and {Wade}, L.~E. and {Wade}, M. and
		  {Waldman}, S.~J. and {Walker}, M. and {Wallace}, L. and
		  {Walsh}, S. and {Wang}, G. and {Wang}, H. and {Wang}, M.
		  and {Wang}, X. and {Wang}, Y. and {Ward}, H. and {Ward},
		  R.~L. and {Warner}, J. and {Was}, M. and {Weaver}, B. and
		  {Wei}, L. -W. and {Weinert}, M. and {Weinstein}, A.~J. and
		  {Weiss}, R. and {Welborn}, T. and {Wen}, L. and
		  {We{\ss}els}, P. and {Westphal}, T. and {Wette}, K. and
		  {Whelan}, J.~T. and {Whitcomb}, S.~E. and {White}, D.~J.
		  and {Whiting}, B.~F. and {Wiesner}, K. and {Wilkinson}, C.
		  and {Willems}, P.~A. and {Williams}, L. and {Williams},
		  R.~D. and {Williamson}, A.~R. and {Willis}, J.~L. and
		  {Willke}, B. and {Wimmer}, M.~H. and {Winkelmann}, L. and
		  {Winkler}, W. and {Wipf}, C.~C. and {Wiseman}, A.~G. and
		  {Wittel}, H. and {Woan}, G. and {Worden}, J. and {Wright},
		  J.~L. and {Wu}, G. and {Yablon}, J. and {Yakushin}, I. and
		  {Yam}, W. and {Yamamoto}, H. and {Yancey}, C.~C. and {Yap},
		  M.~J. and {Yu}, H. and {Yvert}, M. and {Zadro{\.Z}ny}, A.
		  and {Zangrando}, L. and {Zanolin}, M. and {Zendri}, J. -P.
		  and {Zevin}, M. and {Zhang}, F. and {Zhang}, L. and
		  {Zhang}, M. and {Zhang}, Y. and {Zhao}, C. and {Zhou}, M.
		  and {Zhou}, Z. and {Zhu}, X.~J. and {Zucker}, M.~E. and
		  {Zuraw}, S.~E. and {Zweizig}, J. and {LIGO Scientific
		  Collaboration} and {Virgo Collaboration}},
  title		= {{Observation of Gravitational Waves from a Binary Black
		  Hole Merger}},
  journal	= {\prl},
  year		= 2016,
  month		= feb,
  volume	= {116},
  number	= {6},
  eid		= {061102},
  pages		= {061102},
  doi		= {10.1103/PhysRevLett.116.061102},
  archiveprefix	= {arXiv},
  eprint	= {1602.03837},
  primaryclass	= {gr-qc},
  adsurl	= {https://ui.adsabs.harvard.edu/abs/2016PhRvL.116f1102A}
}

@Article{	  2017apj...851l..25f,
  author	= {{Fishbach}, M. and {Holz}, D.~E.},
  title		= "{Where Are LIGO's Big Black Holes?}",
  journal	= {\apjl},
  archiveprefix	= "arXiv",
  eprint	= {1709.08584},
  primaryclass	= "astro-ph.HE",
  year		= 2017,
  month		= dec,
  volume	= 851,
  eid		= {L25},
  pages		= {L25},
  doi		= {10.3847/2041-8213/aa9bf6},
  adsurl	= {http://adsabs.harvard.edu/abs/2017ApJ...851L..25F}
}

@Article{	  2017mnras.471.2801s,
  author	= {{Stevenson}, S. and {Berry}, C.~P.~L. and {Mandel}, I.},
  title		= "{Hierarchical analysis of gravitational-wave measurements
		  of binary black hole spin-orbit misalignments}",
  journal	= {\mnras},
  archiveprefix	= "arXiv",
  eprint	= {1703.06873},
  primaryclass	= "astro-ph.HE",
  year		= 2017,
  month		= nov,
  volume	= 471,
  pages		= {2801-2811},
  doi		= {10.1093/mnras/stx1764},
  adsurl	= {http://adsabs.harvard.edu/abs/2017MNRAS.471.2801S}
}

@Article{	  2019mnras.486.1086m,
  author	= {{Mandel}, Ilya and {Farr}, Will M. and {Gair}, Jonathan
		  R.},
  title		= "{Extracting distribution parameters from multiple
		  uncertain observations with selection biases}",
  journal	= {\mnras},
  year		= 2019,
  month		= jun,
  volume	= {486},
  number	= {1},
  pages		= {1086-1093},
  doi		= {10.1093/mnras/stz896},
  archiveprefix	= {arXiv},
  eprint	= {1809.02063},
  primaryclass	= {physics.data-an},
  adsurl	= {https://ui.adsabs.harvard.edu/abs/2019MNRAS.486.1086M}
}

@Article{	  2019phrvd..99j3015t,
  author	= {{Tsukada}, Leo and {Callister}, Thomas and {Matas}, Andrew
		  and {Meyers}, Patrick},
  title		= "{First search for a stochastic gravitational-wave
		  background from ultralight bosons}",
  journal	= {\prd},
  year		= 2019,
  month		= may,
  volume	= {99},
  number	= {10},
  eid		= {103015},
  pages		= {103015},
  doi		= {10.1103/PhysRevD.99.103015},
  archiveprefix	= {arXiv},
  eprint	= {1812.09622},
  primaryclass	= {astro-ph.HE},
  adsurl	= {https://ui.adsabs.harvard.edu/abs/2019PhRvD..99j3015T}
}

@Article{	  2019phrvd.100d3012w,
  author	= {{Wysocki}, Daniel and {Lange}, Jacob and {O'Shaughnessy},
		  Richard},
  title		= {{Reconstructing phenomenological distributions of compact
		  binaries via gravitational wave observations}},
  journal	= {\prd},
  year		= 2019,
  month		= aug,
  volume	= {100},
  number	= {4},
  eid		= {043012},
  pages		= {043012},
  doi		= {10.1103/PhysRevD.100.043012},
  archiveprefix	= {arXiv},
  eprint	= {1805.06442},
  primaryclass	= {gr-qc},
  adsurl	= {https://ui.adsabs.harvard.edu/abs/2019PhRvD.100d3012W}
}

@Article{	  2020apj...893...35d,
  author	= {{Doctor}, Z. and {Wysocki}, D. and {O'Shaughnessy}, R. and
		  {Holz}, D.~E. and {Farr}, B.},
  title		= "{Black Hole Coagulation: Modeling Hierarchical Mergers in
		  Black Hole Populations}",
  journal	= {\apj},
  year		= 2020,
  month		= apr,
  volume	= {893},
  number	= {1},
  eid		= {35},
  pages		= {35},
  doi		= {10.3847/1538-4357/ab7fac},
  archiveprefix	= {arXiv},
  eprint	= {1911.04424},
  primaryclass	= {astro-ph.HE},
  adsurl	= {https://ui.adsabs.harvard.edu/abs/2020ApJ...893...35D}
}

@Article{2020ApJ...894..129S,
  author	= {{Safarzadeh}, Mohammadtaher and {Farr}, Will M. and
		  {Ramirez-Ruiz}, Enrico},
  title		= "{A Trend in the Effective Spin Distribution of LIGO Binary
		  Black Holes with Mass}",
  journal	= {\apj},
  year		= 2020,
  month		= may,
  volume	= {894},
  number	= {2},
  eid		= {129},
  pages		= {129},
  doi		= {10.3847/1538-4357/ab80be},
  archiveprefix	= {arXiv},
  eprint	= {2001.06490},
  primaryclass	= {gr-qc},
  adsurl	= {https://ui.adsabs.harvard.edu/abs/2020ApJ...894..129S}
}

@Article{	  2020apj...895..128m,
  author	= {{Miller}, Simona and {Callister}, Thomas A. and {Farr},
		  Will M.},
  title		= "{The Low Effective Spin of Binary Black Holes and
		  Implications for Individual Gravitational-wave Events}",
  journal	= {\apj},
  year		= 2020,
  month		= jun,
  volume	= {895},
  number	= {2},
  eid		= {128},
  pages		= {128},
  doi		= {10.3847/1538-4357/ab80c0},
  archiveprefix	= {arXiv},
  eprint	= {2001.06051},
  primaryclass	= {astro-ph.HE},
  adsurl	= {https://ui.adsabs.harvard.edu/abs/2020ApJ...895..128M}
}

@Article{	  2021apj...909l..23e,
  author	= {{Ezquiaga}, Jose Mar{\'\i}a and {Holz}, Daniel E.},
  title		= "{Jumping the Gap: Searching for LIGO's Biggest Black
		  Holes}",
  journal	= {\apjl},
  year		= 2021,
  month		= mar,
  volume	= {909},
  number	= {2},
  eid		= {L23},
  pages		= {L23},
  doi		= {10.3847/2041-8213/abe638},
  adsurl	= {https://ui.adsabs.harvard.edu/abs/2021ApJ...909L..23E}
}

@Article{	  2021apj...913l..19t,
  author	= {{Tiwari}, Vaibhav and {Fairhurst}, Stephen},
  title		= {{The Emergence of Structure in the Binary Black Hole Mass
		  Distribution}},
  journal	= {\apjl},
  year		= 2021,
  month		= jun,
  volume	= {913},
  number	= {2},
  eid		= {L19},
  pages		= {L19},
  doi		= {10.3847/2041-8213/abfbe7},
  archiveprefix	= {arXiv},
  eprint	= {2011.04502},
  primaryclass	= {astro-ph.HE},
  adsurl	= {https://ui.adsabs.harvard.edu/abs/2021ApJ...913L..19T}
}

@Article{	  2021apj...913l..23e,
  author	= {{Edelman}, Bruce and {Doctor}, Zoheyr and {Farr}, Ben},
  title		= "{Poking Holes: Looking for Gaps in LIGO/Virgo's Black Hole
		  Population}",
  journal	= {\apjl},
  year		= 2021,
  month		= jun,
  volume	= {913},
  number	= {2},
  eid		= {L23},
  pages		= {L23},
  doi		= {10.3847/2041-8213/abfdb3},
  archiveprefix	= {arXiv},
  eprint	= {2104.07783},
  primaryclass	= {astro-ph.HE},
  adsurl	= {https://ui.adsabs.harvard.edu/abs/2021ApJ...913L..23E}
}

@Article{	  2021apj...915l..35k,
  author	= {{Kimball}, Chase and {Talbot}, Colm and {Berry},
		  Christopher P.~L. and {Zevin}, Michael and {Thrane}, Eric
		  and {Kalogera}, Vicky and {Buscicchio}, Riccardo and
		  {Carney}, Matthew and {Dent}, Thomas and {Middleton},
		  Hannah and {Payne}, Ethan and {Veitch}, John and
		  {Williams}, Daniel},
  title		= "{Evidence for Hierarchical Black Hole Mergers in the
		  Second LIGO-Virgo Gravitational Wave Catalog}",
  journal	= {\apjl},
  year		= 2021,
  month		= jul,
  volume	= {915},
  number	= {2},
  eid		= {L35},
  pages		= {L35},
  doi		= {10.3847/2041-8213/ac0aef},
  archiveprefix	= {arXiv},
  eprint	= {2011.05332},
  primaryclass	= {astro-ph.HE},
  adsurl	= {https://ui.adsabs.harvard.edu/abs/2021ApJ...915L..35K}
}

@Article{	  2021apj...921l..15g,
  author	= {{Galaudage}, Shanika and {Talbot}, Colm and {Nagar},
		  Tushar and {Jain}, Deepnika and {Thrane}, Eric and
		  {Mandel}, Ilya},
  title		= "{Building Better Spin Models for Merging Binary Black
		  Holes: Evidence for Nonspinning and Rapidly Spinning Nearly
		  Aligned Subpopulations}",
  journal	= {\apjl},
  year		= 2021,
  month		= nov,
  volume	= {921},
  number	= {1},
  eid		= {L15},
  pages		= {L15},
  doi		= {10.3847/2041-8213/ac2f3c},
  archiveprefix	= {arXiv},
  eprint	= {2109.02424},
  primaryclass	= {gr-qc},
  adsurl	= {https://ui.adsabs.harvard.edu/abs/2021ApJ...921L..15G}
}

@Article{	  2021arxiv210409508c,
  author	= {{Callister}, T.~A.},
  title		= "{A Thesaurus for Common Priors in Gravitational-Wave
		  Astronomy}",
  journal	= {arXiv e-prints},
  year		= 2021,
  month		= apr,
  eid		= {arXiv:2104.09508},
  pages		= {arXiv:2104.09508},
  doi		= {10.48550/arXiv.2104.09508},
  archiveprefix	= {arXiv},
  eprint	= {2104.09508},
  primaryclass	= {gr-qc},
  adsurl	= {https://ui.adsabs.harvard.edu/abs/2021arXiv210409508C}
}

@Article{	  2021natas...5..749g,
  author	= {{Gerosa}, Davide and {Fishbach}, Maya},
  title		= "{Hierarchical mergers of stellar-mass black holes and
		  their gravitational-wave signatures}",
  journal	= {Nature Astronomy},
  year		= 2021,
  month		= jul,
  volume	= {5},
  pages		= {749-760},
  doi		= {10.1038/s41550-021-01398-w},
  archiveprefix	= {arXiv},
  eprint	= {2105.03439},
  primaryclass	= {astro-ph.HE},
  adsurl	= {https://ui.adsabs.harvard.edu/abs/2021NatAs...5..749G}
}

@Article{	  2021phrvl.126o1102n,
  author	= {{Ng}, Ken K.~Y. and {Vitale}, Salvatore and {Hannuksela},
		  Otto A. and {Li}, Tjonnie G.~F.},
  title		= "{Constraints on Ultralight Scalar Bosons within Black Hole
		  Spin Measurements from the LIGO-Virgo GWTC-2}",
  journal	= {\prl},
  year		= 2021,
  month		= apr,
  volume	= {126},
  number	= {15},
  eid		= {151102},
  pages		= {151102},
  doi		= {10.1103/PhysRevLett.126.151102},
  archiveprefix	= {arXiv},
  eprint	= {2011.06010},
  primaryclass	= {gr-qc},
  adsurl	= {https://ui.adsabs.harvard.edu/abs/2021PhRvL.126o1102N}
}

@Article{	  2021ptep.2021ea101a,
  author	= {{Akutsu}, T. and {Ando}, M. and {Arai}, K. and {Arai}, Y.
		  and {Araki}, S. and {Araya}, A. and {Aritomi}, N. and
		  {Aso}, Y. and {Bae}, S. and {Bae}, Y. and {Baiotti}, L. and
		  {Bajpai}, R. and {Barton}, M.~A. and {Cannon}, K. and
		  {Capocasa}, E. and {Chan}, M. and {Chen}, C. and {Chen}, K.
		  and {Chen}, Y. and {Chu}, H. and {Chu}, Y. -K. and
		  {Eguchi}, S. and {Enomoto}, Y. and {Flaminio}, R. and
		  {Fujii}, Y. and {Fukunaga}, M. and {Fukushima}, M. and
		  {Ge}, G. and {Hagiwara}, A. and {Haino}, S. and {Hasegawa},
		  K. and {Hayakawa}, H. and {Hayama}, K. and {Himemoto}, Y.
		  and {Hiranuma}, Y. and {Hirata}, N. and {Hirose}, E. and
		  {Hong}, Z. and {Hsieh}, B.~H. and {Huang}, C. -Z. and
		  {Huang}, P. and {Huang}, Y. and {Ikenoue}, B. and {Imam},
		  S. and {Inayoshi}, K. and {Inoue}, Y. and {Ioka}, K. and
		  {Itoh}, Y. and {Izumi}, K. and {Jung}, K. and {Jung}, P.
		  and {Kajita}, T. and {Kamiizumi}, M. and {Kanda}, N. and
		  {Kang}, G. and {Kawaguchi}, K. and {Kawai}, N. and
		  {Kawasaki}, T. and {Kim}, C. and {Kim}, J.~C. and {Kim},
		  W.~S. and {Kim}, Y. -M. and {Kimura}, N. and {Kita}, N. and
		  {Kitazawa}, H. and {Kojima}, Y. and {Kokeyama}, K. and
		  {Komori}, K. and {Kong}, A.~K.~H. and {Kotake}, K. and
		  {Kozakai}, C. and {Kozu}, R. and {Kumar}, R. and {Kume}, J.
		  and {Kuo}, C. and {Kuo}, H. -S. and {Kuroyanagi}, S. and
		  {Kusayanagi}, K. and {Kwak}, K. and {Lee}, H.~K. and {Lee},
		  H.~W. and {Lee}, R. and {Leonardi}, M. and {Lin}, L.~C. -C.
		  and {Lin}, C. -Y. and {Lin}, F. -L. and {Liu}, G.~C. and
		  {Luo}, L. -W. and {Marchio}, M. and {Michimura}, Y. and
		  {Mio}, N. and {Miyakawa}, O. and {Miyamoto}, A. and
		  {Miyazaki}, Y. and {Miyo}, K. and {Miyoki}, S. and
		  {Morisaki}, S. and {Moriwaki}, Y. and {Nagano}, K. and
		  {Nagano}, S. and {Nakamura}, K. and {Nakano}, H. and
		  {Nakano}, M. and {Nakashima}, R. and {Narikawa}, T. and
		  {Negishi}, R. and {Ni}, W. -T. and {Nishizawa}, A. and
		  {Obuchi}, Y. and {Ogaki}, W. and {Oh}, J.~J. and {Oh},
		  S.~H. and {Ohashi}, M. and {Ohishi}, N. and {Ohkawa}, M.
		  and {Okutomi}, K. and {Oohara}, K. and {Ooi}, C.~P. and
		  {Oshino}, S. and {Pan}, K. and {Pang}, H. and {Park}, J.
		  and {Arellano}, F.~E. Pe{\~n}a and {Pinto}, I. and {Sago},
		  N. and {Saito}, S. and {Saito}, Y. and {Sakai}, K. and
		  {Sakai}, Y. and {Sakuno}, Y. and {Sato}, S. and {Sato}, T.
		  and {Sawada}, T. and {Sekiguchi}, T. and {Sekiguchi}, Y.
		  and {Shibagaki}, S. and {Shimizu}, R. and {Shimoda}, T. and
		  {Shimode}, K. and {Shinkai}, H. and {Shishido}, T. and
		  {Shoda}, A. and {Somiya}, K. and {Son}, E.~J. and {Sotani},
		  H. and {Sugimoto}, R. and {Suzuki}, T. and {Suzuki}, T. and
		  {Tagoshi}, H. and {Takahashi}, H. and {Takahashi}, R. and
		  {Takamori}, A. and {Takano}, S. and {Takeda}, H. and
		  {Takeda}, M. and {Tanaka}, H. and {Tanaka}, K. and
		  {Tanaka}, K. and {Tanaka}, T. and {Tanaka}, T. and
		  {Tanioka}, S. and {Tapia San Martin}, E.~N. and {Telada},
		  S. and {Tomaru}, T. and {Tomigami}, Y. and {Tomura}, T. and
		  {Travasso}, F. and {Trozzo}, L. and {Tsang}, T. and
		  {Tsubono}, K. and {Tsuchida}, S. and {Tsuzuki}, T. and
		  {Tuyenbayev}, D. and {Uchikata}, N. and {Uchiyama}, T. and
		  {Ueda}, A. and {Uehara}, T. and {Ueno}, K. and {Ueshima},
		  G. and {Uraguchi}, F. and {Ushiba}, T. and {van Putten},
		  M.~H.~P.~M. and {Vocca}, H. and {Wang}, J. and {Wu}, C. and
		  {Wu}, H. and {Wu}, S. and {Xu}, W. -R. and {Yamada}, T. and
		  {Yamamoto}, K. and {Yamamoto}, K. and {Yamamoto}, T. and
		  {Yokogawa}, K. and {Yokoyama}, J. and {Yokozawa}, T. and
		  {Yoshioka}, T. and {Yuzurihara}, H. and {Zeidler}, S. and
		  {Zhao}, Y. and {Zhu}, Z. -H.},
  title		= {{Overview of KAGRA: Detector design and construction
		  history}},
  journal	= {Progress of Theoretical and Experimental Physics},
  year		= 2021,
  month		= may,
  volume	= {2021},
  number	= {5},
  eid		= {05A101},
  pages		= {05A101},
  doi		= {10.1093/ptep/ptaa125},
  archiveprefix	= {arXiv},
  eprint	= {2005.05574},
  primaryclass	= {physics.ins-det},
  adsurl	= {https://ui.adsabs.harvard.edu/abs/2021PTEP.2021eA101A}
}

@Article{	  2022apj...927..231f,
  author	= {{Fragione}, Giacomo and {Kocsis}, Bence and {Rasio},
		  Frederic A. and {Silk}, Joseph},
  title		= "{Repeated Mergers, Mass-gap Black Holes, and Formation of
		  Intermediate-mass Black Holes in Dense Massive Star
		  Clusters}",
  journal	= {\apj},
  year		= 2022,
  month		= mar,
  volume	= {927},
  number	= {2},
  eid		= {231},
  pages		= {231},
  doi		= {10.3847/1538-4357/ac5026},
  archiveprefix	= {arXiv},
  eprint	= {2107.04639},
  primaryclass	= {astro-ph.GA},
  adsurl	= {https://ui.adsabs.harvard.edu/abs/2022ApJ...927..231F}
}

@Article{2022ApJ...928..155T,
  author	= {{Tiwari}, Vaibhav},
  title		= "{Exploring Features in the Binary Black Hole Population}",
  journal	= {\apj},
  year		= 2022,
  month		= apr,
  volume	= {928},
  number	= {2},
  eid		= {155},
  pages		= {155},
  doi		= {10.3847/1538-4357/ac589a},
  archiveprefix	= {arXiv},
  eprint	= {2111.13991},
  primaryclass	= {astro-ph.HE},
  adsurl	= {https://ui.adsabs.harvard.edu/abs/2022ApJ...928..155T}
}

@Article{	  2022apj...931..108f,
  author	= {{Farah}, Amanda and {Fishbach}, Maya and {Essick}, Reed
		  and {Holz}, Daniel E. and {Galaudage}, Shanika},
  title		= "{Bridging the Gap: Categorizing Gravitational-wave Events
		  at the Transition between Neutron Stars and Black Holes}",
  journal	= {\apj},
  year		= 2022,
  month		= jun,
  volume	= {931},
  number	= {2},
  eid		= {108},
  pages		= {108},
  doi		= {10.3847/1538-4357/ac5f03},
  archiveprefix	= {arXiv},
  eprint	= {2111.03498},
  primaryclass	= {astro-ph.HE},
  adsurl	= {https://ui.adsabs.harvard.edu/abs/2022ApJ...931..108F}
}

@Article{	  2022lrr....25....1m,
  author	= {{Mandel}, Ilya and {Broekgaarden}, Floor S.},
  title		= "{Rates of compact object coalescences}",
  journal	= {Living Reviews in Relativity},
  year		= 2022,
  month		= dec,
  volume	= {25},
  number	= {1},
  eid		= {1},
  pages		= {1},
  doi		= {10.1007/s41114-021-00034-3},
  archiveprefix	= {arXiv},
  eprint	= {2107.14239},
  primaryclass	= {astro-ph.HE},
  adsurl	= {https://ui.adsabs.harvard.edu/abs/2022LRR....25....1M}
}

@Article{	  2022mnras.509.5454r,
  author	= {{Rinaldi}, Stefano and {Del Pozzo}, Walter},
  title		= "{(H)DPGMM: a hierarchy of Dirichlet process Gaussian
		  mixture models for the inference of the black hole mass
		  function}",
  journal	= {\mnras},
  year		= 2022,
  month		= feb,
  volume	= {509},
  number	= {4},
  pages		= {5454-5466},
  doi		= {10.1093/mnras/stab3224},
  archiveprefix	= {arXiv},
  eprint	= {2109.05960},
  primaryclass	= {astro-ph.IM},
  adsurl	= {https://ui.adsabs.harvard.edu/abs/2022MNRAS.509.5454R}
}

@Article{	  2022phrvd.106b3020y,
  author	= {{Yuan}, Chen and {Jiang}, Yang and {Huang}, Qing-Guo},
  title		= "{Constraints on an ultralight scalar boson from Advanced
		  LIGO and Advanced Virgo's first three observing runs using
		  the stochastic gravitational-wave background}",
  journal	= {\prd},
  year		= 2022,
  month		= jul,
  volume	= {106},
  number	= {2},
  eid		= {023020},
  pages		= {023020},
  doi		= {10.1103/PhysRevD.106.023020},
  archiveprefix	= {arXiv},
  eprint	= {2204.03482},
  primaryclass	= {astro-ph.CO},
  adsurl	= {https://ui.adsabs.harvard.edu/abs/2022PhRvD.106b3020Y}
}

@Article{	  2022phrvd.106j3013m,
  author	= {{Mould}, Matthew and {Gerosa}, Davide and {Taylor},
		  Stephen R.},
  title		= "{Deep learning and Bayesian inference of
		  gravitational-wave populations: Hierarchical black-hole
		  mergers}",
  journal	= {\prd},
  year		= 2022,
  month		= nov,
  volume	= {106},
  number	= {10},
  eid		= {103013},
  pages		= {103013},
  doi		= {10.1103/PhysRevD.106.103013},
  archiveprefix	= {arXiv},
  eprint	= {2203.03651},
  primaryclass	= {astro-ph.HE},
  adsurl	= {https://ui.adsabs.harvard.edu/abs/2022PhRvD.106j3013M}
}

@Article{	  2022phrvl.129f1102e,
  author	= {{Ezquiaga}, Jose Mar{\'\i}a and {Holz}, Daniel E.},
  title		= "{Spectral Sirens: Cosmology from the Full Mass
		  Distribution of Compact Binaries}",
  journal	= {\prl},
  year		= 2022,
  month		= aug,
  volume	= {129},
  number	= {6},
  eid		= {061102},
  pages		= {061102},
  doi		= {10.1103/PhysRevLett.129.061102},
  archiveprefix	= {arXiv},
  eprint	= {2202.08240},
  primaryclass	= {astro-ph.CO},
  adsurl	= {https://ui.adsabs.harvard.edu/abs/2022PhRvL.129f1102E}
}

@Article{	  2024apj...966l..16p,
  author	= {{Payne}, Ethan and {Kremer}, Kyle and {Zevin}, Michael},
  title		= "{Spin Doctors: How to Diagnose a Hierarchical Merger
		  Origin}",
  journal	= {\apjl},
  year		= 2024,
  month		= may,
  volume	= {966},
  number	= {1},
  eid		= {L16},
  pages		= {L16},
  doi		= {10.3847/2041-8213/ad3e82},
  archiveprefix	= {arXiv},
  eprint	= {2402.15066},
  primaryclass	= {gr-qc},
  adsurl	= {https://ui.adsabs.harvard.edu/abs/2024ApJ...966L..16P}
}
\end{document}